\documentclass[preprint,11pt,fleqn,3p]{elsarticle}

\usepackage{amssymb}

\usepackage{lineno}

\usepackage{hyperref}
\hypersetup{
    colorlinks=true,
    linkcolor=blue,
    filecolor=magenta,      
    urlcolor=cyan,
}
\usepackage{setspace}
\usepackage{geometry}
\usepackage{lipsum}
\usepackage{multicol}
\usepackage{balance}       
\usepackage{graphicx}      
\usepackage{svg}
\usepackage{xcolor}
\usepackage{changepage}
\usepackage{caption}    
\usepackage{subfig} 
\usepackage[export]{adjustbox} 
\usepackage{multicol}
\usepackage{bm}
\usepackage{amsmath, array,tabstackengine}
\TABstackMath
\TABstackMathstyle{\displaystyle}
\usepackage{amssymb, mathtools}
\usepackage{pdfpages}
\usepackage{verbatim}  
\usepackage{vector}  
\usepackage{indentfirst}
\usepackage{nicematrix}
\usepackage{caption}
\usepackage{minitoc}
\usepackage{multirow}
\usepackage{floatrow}
\usepackage{enumitem}
\usepackage[nameinlink]{cleveref}
\usepackage{titletoc}
\usepackage{float}
\hypersetup{urlcolor=blue, colorlinks=true, citecolor = blue}  
\hypersetup{pdfauthor={Name}}
\usepackage{cleveref}
\bibpunct{\textcolor{blue}{[}}{\textcolor{blue}{]}}{,}{a}{,}{,}
\usepackage{siunitx}
\usepackage{etoolbox}
\let\oldequation\equation
\let\oldendequation\endequation

\renewenvironment{equation}
  {\linenomathNonumbers\oldequation}
  {\oldendequation\endlinenomath}

\begin{document}
\begin{frontmatter}
\title{A DEM-driven machine learning framework for abrasive wear prediction}
\author[inst1]{Prassana Chandan\corref{equal}}
\affiliation[inst1]{organization={Mechanics of Materials Lab, Department of Mechanical Engineering},
            addressline={Indian Institute of Technology Madras}, 
            city={Chennai},
            postcode={600036}, 
            state={Tamil Nadu},
            country={India}
            }
\author[inst1]{Amiya Prakash Das\corref{equal}}
\author[inst1]{Shakti Swaroop Choudhury}
\author[inst1,inst2]{Ratna Kumar Annabattula\corref{cor}}
\ead{ratna@iitm.ac.in}
\affiliation[inst2]{organization={Center for Soft and Biological Matter},
            addressline={Indian Institute of Technology Madras}, 
            city={Chennai},
            postcode={600036}, 
            state={Tamil Nadu},
            country={India}
            }
\cortext[cor]{Corresponding author}
\cortext[equal]{Equal contribution}

\begin{abstract}
Particle-induced wear is a critical concern in bulk material handling systems, where abrasive interactions accelerate equipment degradation, increase maintenance needs, and raise operational costs. The Discrete Element Method (DEM) and Archard’s wear model are widely adopted for predicting particle-surface wear processes. However, DEM is computationally prohibitive for real-time design and predictive maintenance, often requiring hours to days for a single parametric analysis. We propose a DEM–machine learning (ML) framework to address this limitation that combines physics-based simulations with data-driven efficiency. A dataset of 200 DEM simulations is generated by systematically varying particle size, material, and contacting plate geometric parameters. A few ML models\textemdash linear regression, Lasso and Ridge regularization, decision trees, and a genetic algorithm–optimized artificial neural network (GA-ANN) were trained and evaluated. Feature selection revealed that Archard’s wear constant, particle size, plate angle, and impingement velocity are the dominant predictors of wear. While linear models offered interpretability, their accuracy was limited. The GA-ANN achieved the highest performance (\(R^2 = 0.91\)), effectively capturing nonlinear wear dynamics while reducing computational cost by orders of magnitude. This study demonstrates that physics-informed ML provides a scalable pathway for accurate, real-time wear prediction, enabling predictive maintenance and optimized design in bulk material handling industries.
\end{abstract}


\end{frontmatter}
\section{Introduction}
\noindent Particle-induced wear in bulk material handling equipment\textemdash like chutes, conveyors, and excavators\textemdash poses a critical challenge in industries like mining and construction, leading to frequent maintenance interventions and substantial economic losses~\citep{carter1980mechanism, Hutchings1992,owen2009prediction,che2025novel}. Abrasive wear, caused by intense particle-surface interactions, results in unplanned downtime, financial overheads, and operational unreliability~\citep{suh1980effect,Johanson1982,owen2009prediction}. Accurate wear prediction is vital for improving the life cycles and performance of heavy machinery in bulk handling environments\textemdash minimizing downtime and operational costs~\citep{graff2010discrete,xia2019discrete,Amadi2024}.

The Discrete Element Method (DEM) has emerged as a sought-after numerical tool for analyzing and predicting the particle-induced wear ~\citep{fillot2007modelling,jerier2012normal,pozzetti2018numerical,fransen2021application,thompson2022effect}, yet significant computational challenges remain in translating the simulation results into practical, real-time design and maintenance solutions. Traditional DEM uses Archard's wear equation to model particle-induced wear. It is computationally prohibitive\textemdash requiring several hours to days for a single parametric sweep~\citep{Archard1953,owen2009prediction,jayasundara2022predicting,wang2023parameter,yan2023modelling}. The excessive computational cost limits their applicability in real-time design iterations and predictive maintenance, highlighting the need for innovative solutions to enhance prediction efficiency with accuracy~\citep{thompson2022effect,liskiewicz2023advances,Zhang2024}. 

Recent developments in machine learning have shown promise in addressing computational bottlenecks across various engineering domains~\citep{yan2015discrete,fransen2021application,wallin2022data,irazabal2023methodology,jin2023recent,sose2023review,che2025novel,wang2025machine}.~\citet{Rajput2023} have demonstrated the feasibility of integrating machine learning with DEM to circumvent the limitations of traditional wear prediction strategies. For instance, tree-based models like XGBoost have shown high accuracy in predicting tool wear, highlighting the applicability of machine learning in tribological studies~\citep{Fathi2024,che2025novel}. Furthermore, genetic algorithm-optimized artificial neural networks (GA-ANNs) have successfully captured the nonlinear dynamics inherent in wear phenomena, outperforming classical statistical models~\citep{jin2023recent,Danish2024,hussain2024machine}. However, despite recent advances in wear prediction and damage assessment, scientific machine learning for wear prediction in industrial applications remains largely unexplored, with limited work focusing on comprehensive frameworks that integrate DEM simulations with diverse machine learning models for particle-based wear in bulk handling equipment~\citep{yan2015discrete,fransen2021application,Hasan2022,irazabal2023methodology,Zhu2024}.

This study addresses the fundamental research question: Can machine learning algorithms trained on DEM-generated datasets provide computationally efficient and accurate predictions of wear rates while maintaining the high-fidelity of traditional DEM approaches? Specifically, we aim to develop a machine learning (ML) framework that reduces computational time by orders of magnitude\textemdash preserving predictive accuracy. Furthermore, systematic feature analysis determines the key physical parameters that govern wear prediction accuracy.

The framework directly addresses the bottlenecks hindering large-scale industrial adoption of DEM by leveraging ML models. The framework is trained on a physics-informed dataset comprising 200 systematically varied DEM simulations generated using Altair EDEM~\citep{Engineering2023}. A few ML algorithms, including linear regression, decision trees, and GA-optimized ANNs, are trained and evaluated using this physics-informed dataset. The key contributions of this work are threefold. First, we demonstrate that ML can achieve comparable accuracy to traditional DEM approaches with considerably reduced computational cost and time. Second, we identify key predictors such as impact velocity and particle size through systematic feature importance analysis, providing actionable insights for equipment design optimization~\citep{owen2009prediction,lommen2019particle,Deshpande2024}. Third, we establish a scalable framework that can be readily extended to diverse material systems, bridging the gap between computational tribology research and industrial practice~\citep{xia2019discrete,wallin2022data,Yan2023,wang2025machine}.

The paper is organized as follows:~\Cref{sec:methodology} outlines the methodology, including the DEM simulation framework, Archard's wear calculation, and the suite of machine learning models employed.~\Cref{simulation_details} details the simulation setup, parameter space definition, and data preprocessing steps.~\Cref{resultsAndDiscussions} presents and analyzes the results, comparing the predictive performance of various models and identifying dominant wear-influencing features. Finally,~\Cref{summaryAndConclusion} summarizes the key findings and suggests directions for future research.

\section{Methodology}\label{sec:methodology}
\noindent This study integrates DEM-based simulations with a suite of ML algorithms to predict particle-induced wear on a predefined geometry. The framework comprises simulation-driven data generation, feature engineering, dimensionality reduction, and the development and evaluation of regression models, culminating in deploying a GA-ANN.
\subsection{Discrete element method}
\noindent DEM simulations are conducted using a commercial DEM package \texttt{Altair EDEM} to generate a comprehensive wear dataset across a range of material parameters and configurations~\citep{coetzee2017calibration}. The elastic contact forces are calculated using the Hertz-Mindlin model, which assumes small deformation at particle contact points such that the normal overlap $\delta_{ij}^\text{n}$ remains significantly smaller than \(\min(r_i,r_j)\)~\citep{mindlin1953elastic}. Following the formulations of~\citet{johnson1987contact} and~\citet{thornton2011investigation}, the normal contact force component $F_{ij}^{\text{n}}$ is expressed as
\begin{equation}
    F_{ij}^{\text{n}}=\ -k_{ij}^{\text{n}}\left(\delta_{ij}^\text{n}\right)^{\frac{3}{2}},
\label{eq1}
\end{equation}

\noindent where $k_{ij}^{\text{n}}$ is the normal contact stiffness and $\delta_{ij}^\text{n}$ denotes 
the normal overlap. The tangential contact force is updated incrementally at each time step $\Delta t$ as  
\begin{equation}
    \left(F_{ij}^{\text{s}}\right)^{t} 
    = \left(F_{ij}^{\text{s}}\right)^{t-\Delta t} 
    + k_{ij}^{\text{s}} \, \Delta u_\text{s}^{t},
\label{eq2}
\end{equation}

\noindent where $k_{ij}^{\text{s}}$ is the tangential contact stiffness and $\Delta u_\text{s}^{t}$ 
is the tangential (shear) displacement increment over the time step.

The total inter-particle contact force \(F_{ij}^\text{c}\) combines normal \((F_{ij}^{\text{n}})\) and tangential components \((F_{ij}^{\text{s}})\) vectorially. Particle dynamics are governed by Newton's second law, accounting for contact interactions and gravitational effects to determine particle acceleration. Velocity and position are subsequently obtained through Euler integration of the acceleration field. The tangential force evolution follows an incremental approach based on relative tangential displacement, with sliding triggered when the tangential force exceeds the Coulomb friction criterion, \(F_{ij}^\text{s} \leq \mu F_{ij}^\text{n}\), where $\mu$ is the coefficient of sliding friction between the particles $i$ and $j$.

\subsection{Archard's wear computation}
\noindent Archard's wear model is employed to quantify the wear volume \(V\) at individual contact points across various material parameter combinations and geometric configurations~\citep{Archard1953,Hutchings1992,jayasundara2022predicting}. The wear volume is calculated as:
\begin{equation}
V = W F^\text{n} s,
\label{eq3}
\end{equation}

\noindent where, \(W\) represents the Archard's wear constant (\si{\per\pascal}), \(F^\text{n}\) denotes the normal contact force (\si{\newton}), \(s\) (\si{\meter}) denotes the cumulative sliding distance. For computational implementation, wear is conventionally expressed as geometric changes in surface facet elements, specifically as wear depth rather than volumetric loss. Under the assumption of uniform wear distribution across the contact area $A$, the corresponding wear depth $h$ is determined using, 
\begin{equation}
    h = \dfrac{V}{A} = \dfrac{W F^\text{n} s}{A}.
    \label{eq4}
\end{equation}
\noindent This depth-based representation facilitates visualization of wear patterns and export of spatial wear profiles for further analysis~\citep{jayasundara2022predicting,Zhang2024}.

The Archard's wear model has been extensively validated for DEM simulations by comparing experimental data across various industrial scenarios.~\citet{rojas2019case,jayasundara2022predicting} demonstrated a strong correlation between DEM-predicted wear distributions using Archard's formulation and experimentally observed wear patterns in controlled chute liner tests, establishing the model's predictive capability for industrial equipment. This validation has been corroborated by~\citet{forsstrom2016calibration} and~\citet{chen2017sensitivity}, who reported a good agreement between the predictions and experimental measurements across different material systems and geometrical configurations. These comprehensive validations establish Archard's model as a reliable framework for wear prediction for engineering applications.

In each DEM simulation, the particle impacts on a flat plate geometry are analyzed by systematically varying parameters such as particle impinging velocity, material modulus, friction coefficient, and contact angles. A structured simulation protocol is employed to ensure robust data collection.

\subsection{Machine learning techniques}
\noindent This study implements several supervised machine learning techniques to model the relationship between simulated features and wear outcomes. Each algorithm is briefly described below.

\subsubsection{Linear regression}
\noindent Linear regression models the relationship between a continuous target variable and one or more predictors~\citep{aalen1989linear}. The model is expressed as:
\begin{equation}
    \hat{y}_i = \beta_0 + \sum_{j=1}^p \beta_j x_{ij},
    \label{eq5}
\end{equation}

\noindent where, \(\beta_0\) is the intercept, \(\beta_j\) are the regression coefficients, \(x_{ij}\) are feature values, and \(\hat{y}_i\) is the predicted wear for the \(i\)-th observation. These coefficients are estimated using ordinary least squares (OLS). It minimizes the sum of squared residuals between observed and predicted values, making it the best linear unbiased estimator under the Gauss-Markov theorem~\citep{montgomery2012introduction}. OLS is widely used in engineering and statistics for its simplicity and interoperability. Here, the linear regression served as an interpretable baseline, allowing us to quantify how each physical and geometric input feature contributes to wear prediction without introducing nonlinear complexity.

\subsubsection{Regularization: Lasso (L1) and Ridge (L2)}
\noindent Linear regression models overfit and influence the correlated predictors. To avoid such artifacts, two regularization methods\textemdash Lasso (L1) and Ridge (L2), are applied in conjunction with the linear regression model. Lasso regression adds an L1 penalty term proportional to the absolute values of the coefficients. Its objective function is given as:
\begin{equation*}
    \frac{1}{n}\sum_{i=1}^n (y_i - \hat{y}_i)^2 + \lambda \sum_{j=1}^p \|\beta_j\|.
\label{eq6}
\end{equation*}
The L1 penalty promotes sparsity by forcing some coefficients to precisely zero. This approach allows for selecting a smaller subset of features for analysis~\citep{tibshirani1996regression}. One of the key benefits of using Lasso is its ability to simplify models by removing redundant features, which enhances interpretability and improves generalization~\citep{santosa1986linear}. Ridge regression introduces an L2 penalty, expressed as,
\begin{equation*}
    \frac{1}{k}\sum_{i=1}^k (y_i - \hat{y}_i)^2 + \lambda \sum_{j=1}^p \beta_j^2.
\label{eq7}
\end{equation*}
This shrinks coefficients toward zero without eliminating any predictors, stabilizing estimation when predictors are highly correlated~\citep{hayashi1998data,hoerl1970ridge}.
These methods help eliminate multicollinearity, where predictor variables are highly interdependent, inflating coefficient variances and reducing model reliability. Ridge regression effectively handles multicollinearity by shrinking all coefficients, leading to more stable estimations~\citep{herawati2024performance}. Lasso tends to select one variable from each group of correlated predictors while suppressing others~\citep{santosa1986linear}. In contrast, Ridge distributes shrinkage across all predictors, leading to lower mean squared error when multicollinearity is severe~\citep{herawati2024performance}.

The Lasso model retained only the most influential features, like Archard's constant, particle size, plate angle, and particle's \(Y\) (impinging) velocity, simplifying the feature set without losing substantial predictive power. In contrast, the Ridge model maintained all features but reduced the magnitude of less significant ones, offering robust generalization without discarding potentially useful variables. This combination of Lasso and Ridge provides stability to the wear prediction framework.

\subsubsection{Principal component analysis (PCA)}
\noindent PCA is an unsupervised technique that reduces dimensionality and identifies the principal directions of variance in the data~\citep{JolliffeCadima2016}.~Given a mean-centered data matrix \(X\), PCA computes orthonormal vectors \(w_k\) (the principal components) such that projections \(t_k = X w_k\) capture maximal variance in decreasing order~\citep{JolliffeCadima2016}. These components are ranked by eigenvalues derived from the covariance matrix, allowing for effective dimensionality reduction while preserving the variation in the dataset~\citep{JolliffeCadima2016}.

PCA helps identify clusters of correlated variables by grouping features that load heavily on the same principal components~\citep{arguelles2014new,JolliffeCadima2016}. This grouping informs feature selection by highlighting redundancies; features with low variance contributions or high mutual correlation can be deprioritized or removed~\citep{ibrahim2021feature}. Here, PCA offered valuable insights into the feature structure and supported the validation of results from regularization techniques. Although PCA is not directly used for regression input since it does not consider the target variable, it revealed that elasticity-related variables, like stiffness parameters, tend to cluster together. 

\subsubsection{Decision tree regressor}
\noindent Decision tree regression is a non-parametric supervised learning method that predicts continuous outcomes by creating a hierarchical model of decision rules based on feature thresholds. The data is recursively split at each node into subsets to minimize prediction error, using algorithms like Classification and Regression Trees~\citep{Breiman1984}. The leaf nodes represent regions where final predictions are computed as the mean outcome of training samples within that region. This piecewise-constant structure makes decision trees highly interpretable, as decision paths can be easily visualized and traced for any prediction~\citep{Breiman1984,Bohemke2020}.

Decision trees are susceptible to overfitting, especially when deep, as they may capture training data peculiarities instead of general patterns, which reduces generalization to new data~\citep{Loh2014}. To mitigate this, hyperparameter tuning and pruning techniques are essential. Pre-pruning limits growth based on depth or impurity, while post-pruning removes non-informative branches~\citep{Bohemke2020}. While interpretable, decision trees can be unstable; small changes in training data can lead to drastically different structures~\citep{Loh2014}. This instability can diminish performance on unseen data, often necessitating regularization or ensemble methods like Random Forests for improved robustness~\citep{ali2012random}.

Here, we used a shallow decision tree (depth 1–2) to illustrate simple splits in wear-driving features. It provided high interpretability but limited predictive performance, leading to the adoption of more expressive models like neural networks in later stages.

\subsubsection{Genetic algorithm–optimized artificial neural network (GA‑ANN)}
\noindent Artificial Neural Networks (ANNs) consist of interconnected layers of neurons, where inputs are transformed into outputs via weighted sums and activation functions, such as ReLU or sigmoid, to learn complex relationships between inputs and targets~\citep{Goodfellow2016Deep}. To design an effective ANN for wear rate prediction, a Genetic Algorithm (GA) is employed to optimize key hyperparameters like the number of hidden layers, neuron counts per layer, and initial weight configurations by encoding them as chromosomes and evolving a population across multiple generations~\citep{Yao1999Evolving,Nikbakht2021Optimizing}. The GA uses operators like selection, crossover, and mutation to navigate the search space, effectively exploring architecture and weight configurations to enhance model performance~\citep{Yao1999Evolving,Xiao2020Efficient}.

The aforementioned GA-ANN hybrid framework improves over conventional gradient-based training by avoiding issues such as local minima and sensitivity to weight initialization. It provides a global search mechanism that explores varied network configurations and identifies top-performing solutions even in complex, non-convex problem spaces~\citep{Nikbakht2021Optimizing,Loh2014}. In this study, the GA runs for 25 generations, evolving network structure and weight initializations, with \(R^2\) used as the fitness function. The optimal architecture consisted of four hidden layers with neuron counts: [449, 24, 500, 385]. This configuration reflects how GA can discover non-intuitive architectures that traditional tuning might miss.

\section{Details of simulations}\label{simulation_details}
\subsection{DEM simulation setup for wear analysis}
\begin{figure}[H]
    \centering
    \includegraphics[width=0.8\linewidth]{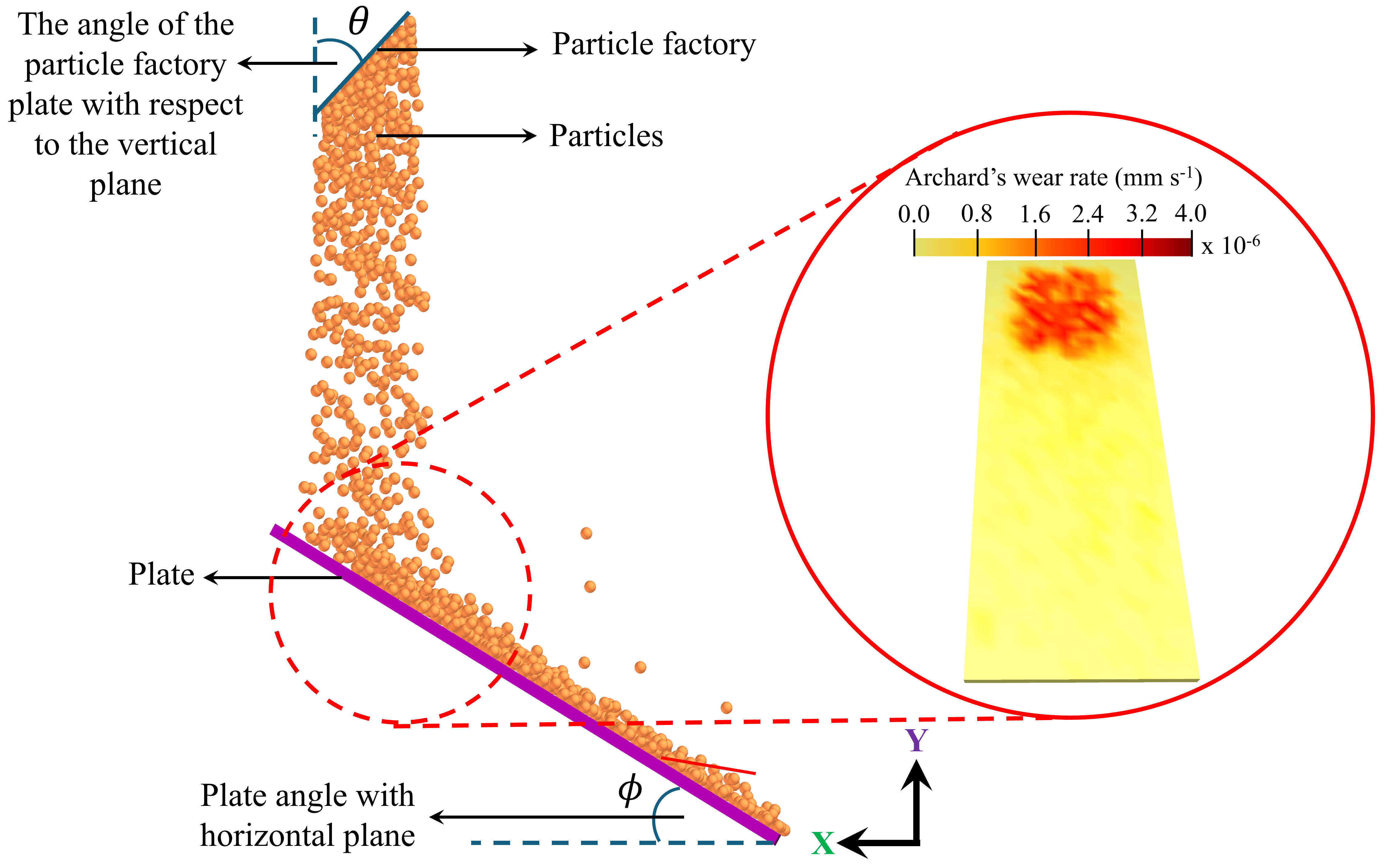}
    \caption{Schematic representation of the DEM simulation domain for data collection. The callout highlights Archard's wear pattern on the plate, with a color gradient indicating wear intensity from low (yellow colour) to high (red colour).}
    \label{fig:simulation_setup}
\end{figure}

\noindent A three-dimensional DEM solver (EDEM) simulates particle-induced wear on inclined surfaces, see~\Cref{fig:simulation_setup}. The setup features a planar plate and a particle factory, oriented at angles \(\phi\) and \(\theta\), respectively, allowing for a systematic investigation of how impact angles affect wear depths/rate. Monodisperse particles are used in simulations with a constant generation rate. Material properties for particles and surfaces are defined using representative material parameters for materials like coal and steel, while particle-contact surface interactions are modeled with the Hertz-Mindlin contact model~\citep{mindlin1953elastic}. Gravitational force and the contact forces govern the particle motion, and wear is quantified using Archard's wear model,~\Cref{eq3}. The material parameters are detailed in ~\Cref{tab:simulation_parameters_and_their_ranges}. A fixed time step of \(\Delta t \approx 20\%\) of the Rayleigh time step ensures stable computations. The computational domain is structured to ensure sufficient particle generation and flow rate.

The Archard's wear predictions from the DEM simulation are verified through comparison with both experimental data and DEM results for a screw conveyor reported by \citet{yang2021screwwear}. Following their setup, the DEM model parameters are defined as: particle radius \(r_p = \SI{8}{\milli\meter}\), particle density \(\rho_p = \SI{1500}{\kilogram\per\meter\cubed}\) and Young's modulus \(\SI{1}{\giga\pascal}\), friction coefficient of 0.56, and restitution coefficient of 0.35 for particle\textendash particle interactions. For particle\textendash plate interactions, the friction and restitution coefficients are set to 0.46 and 0.50, respectively. The plate properties are specified as density \(\SI{7800}{\kilogram\per\meter\cubed}\) and Young’s modulus \(\SI{210}{\giga\pascal}\).
The Archard's wear constant is fixed at \(1 \times 10^{-11}\, \si{\per\pascal}\), consistent with the experimental study. Analysis of particle\textendash plate interactions indicates that the predicted Archard's wear depth is on the order of \(10^{-6}\, \si{\milli\meter\per\second}\), which shows good agreement with both the experimental observations and the reference DEM simulations, see zoomed section for Archard's wear rate in~\Cref{fig:simulation_setup}.

For the training and testing datasets, multiple simulations are conducted with varying \(\phi\), \(\theta\), particle properties, and interaction parameters. Each configuration allowed the particle flow to be steady before collecting wear data. The simulation tracked key parameters, including particle positions, velocities, contact forces, and accumulated wear depth. Data is exported regularly for time-series analysis of wear progression and pattern formation. This methodical approach produced a comprehensive dataset for subsequent analysis.

\subsection{Data curation}
\noindent The parameter space for the simulations is carefully identified to encompass a range of material parameters and geometrical configurations relevant to industrial bulk material handling, as delineated below.
\begin{itemize}[labelindent=1em, labelsep=0.5cm, leftmargin=*]
    \item \textbf{Material parameters}: Young's modulus, Poisson's ratio, and density for particles and equipment surfaces.
    \item \textbf{Interaction parameters}: Coefficients of restitution, static friction, and rolling friction.
    \item \textbf{Geometric configuration}: Plate angle, factory angle, particle size.
    \item \textbf{Kinematic parameters}: Initial particle impinging velocity.
    \item \textbf{Derived parameters}: Effective elastic and shear moduli, damping ratio, and damping coefficient.
\end{itemize}

Each parameter is assigned a range of values based on material handbooks, industry standards, and prior research in bulk material handling. A uniform distribution sampling approach, rather than a structured grid, ensures comprehensive parameter space coverage. This process provides a better space-filling property for the subsequent ML tasks. A custom Python script is developed to generate the simulation input dataset. Parameter ranges and constraints are defined to ensure realistic input values, followed by uniform random sampling within these ranges to create diverse samples. Each combination is verified for physical consistency, resulting in 200 unique parameter sets, with 80\% allocated for model training and 20\% for testing and validation.
\begin{table}[H]
\centering
\caption{List of the mapping feature indices to the physical parameters. The subscripts \(\text{p}\) and \(\text{e}\) refer to the particle and equipment, respectively.}
\begin{tabular}{ll|ll}
\hline \hline
\textbf{Symbol} & \textbf{Feature Name} &\textbf{Symbol} & \textbf{Feature Name}\\
\hline \hline
\(X_1\)  & Young's Modulus \((E_\text{p})\)  &\(X_2\)  & Shear Modulus \((G_\text{p})\) \\
\(X_3\)  & Poisson's Ratio \((\nu_\text{p})\) &\(X_4\)  & Density \(\rho_\text{p}\)  \\
\(X_5\)  & Young's Modulus \((E_\text{e})\) &\(X_6\)  & Shear Modulus \((G_\text{e})\) \\
\(X_7\)  & Poisson's Ratio \((\nu_\text{e})\)  &\(X_8\)  & Density \((\rho_\text{e})\) \\
\(X_9\)  & Coefficient of Restitution &\(X_{10}\) & Coefficient of Rolling Friction \\
\(X_{11}\) & Coefficient of Static Friction & \(X_{12}\) & Archard's Wear Constant \\
\(X_{13}\) & Plate Angle & \(X_{14}\) & Factory Angle \\
\(X_{15}\) & Particle Size & \(X_{16}\) & Particle \(Y\) Velocity \\
\(X_{17}\) & Effective Young's Modulus & \(X_{18}\) & Effective Shear Modulus \\
\(X_{19}\) & Damping Ratio &\(X_{20}\) & Damping Coefficient \\
\hline \hline
\end{tabular}
\label{tab:feature_mapping}
\end{table}

\begin{table}[H]
\centering
\caption{Mapping of feature indices to physical parameters with value ranges}
\begin{tabular}{lcc}
\hline \hline
\textbf{Feature Name} & \textbf{Lower Limit} & \textbf{Upper Limit} \\
\hline \hline
Shear Modulus (particle) [\si{\giga\pascal}] & $2.0$ & $10.0$ \\
Poisson's Ratio (particle) [-] & 0.2 & 0.35 \\
Density (particle) [\si{\kilogram\per\meter\cubed}] & 2000 & 4000 \\
Shear Modulus (Equipment) [\si{\giga\pascal}] & $10.0$ & $100.0$ \\
Poisson's Ratio (Equipment) [-] & 0.2 & 0.35 \\
Density (Equipment) [\si{\kilogram\per\meter\cubed}] & 6000 & 8000 \\
Coefficient of Restitution (Interaction) [-]& 0.2 & 0.8 \\
Coefficient of Rolling Friction (Interaction) [-] & 0.01 & 0.1 \\
Coefficient of Static Friction (Interaction) [-]& 0.4 & 0.55 \\
Archard's Wear Constant [\si{\per\pascal}] & $1.0\times10^{-10}$ & $1.0\times10^{-8}$ \\
Plate Angle [\SI{}{\degree}]  & 0 & 70 \\
Factory Angle [\SI{}{\degree}]  & -90 & 0 \\
Particle Size [\si{\milli\meter}] & 5 & 40 \\
Particle \(Y\) Velocity [\si{\meter\per\second}] & -10  & 0 \\
\hline \hline
\end{tabular}
\label{tab:simulation_parameters_and_their_ranges}
\end{table}

The generated parameter sets are transformed into simulation configurations using \texttt{EDEMpy}, allowing batch processing. A base simulation template is created for each parameter set, ensuring consistent file management. Geometry and material properties are configured, and the Hertz-Mindlin contact model is implemented. The \texttt{Deck} class in \texttt{EDEMpy} facilitated precise control over simulation parameters.

\subsection{Data processing and feature extraction}
\noindent Data processing involved extracting time-series features like wear depth, time variation of cumulative wear, and fitting a linear regression model to compute the wear rate slope. The \(R^2\) metric, exceeding 0.95, confirmed the linear wear rate assumption. These wear rates became target outputs for subsequent ML models. The consistency of linear trends validated a single-slope representation, simplifying analysis and reducing noise sensitivity while ensuring compatibility with various model training workflows.

\subsubsection{Feature compilation and preprocessing}
\noindent The complete dataset is constructed by merging simulation parameters with calculated wear rates. In addition to the directly measured parameters, several derived features are incorporated to capture the mechanical interactions governing wear. These features are defined as follows:

\begin{equation}
    \dfrac{1}{E^{*}} = \dfrac{1 - \nu_1^2}{E_1} + \dfrac{1 - \nu_2^2}{E_2},
\end{equation}
\begin{equation}
    \dfrac{1}{G^{*}} = \dfrac{2 - \nu_1}{G_1} + \dfrac{2 - \nu_2}{G_2},
\end{equation}
\begin{equation}
    \xi = \dfrac{\ln(e)}{\sqrt{\ln^2(e) + \pi^2}},
\end{equation}
\begin{equation}
    \eta = \dfrac{-2\,\sqrt{m\,k^\text{n}}\;\ln(e)}{\sqrt{\ln^2(e) + \pi^2}},
\end{equation}

\noindent where \(E^*\) is the effective Young’s modulus, which governs the normal contact force \(F^\text{n}\) and is directly proportional to the normal wear component in Archard’s model. Similarly,
\(G^*\) denotes the effective shear modulus, governing the tangential contact force \(F^\text{s}\), which is linked to sliding distance (or tangential wear). The parameter \(\xi\) is the damping ratio, while \(\eta\) is the damping coefficient; both are derived from the coefficient of restitution \(e\) and quantify energy dissipation during particle impacts. Here, \(m = m_{1}m_{2}/m_{1}+m_{2}\) is the reduced mass of the colliding pair. These derived parameters are chosen because they maintain a direct physical relationship with the wear process, allowing simpler models like linear regression or decision trees to capture their effects efficiently. By incorporating them, the feature set provides mechanistic insights influencing wear, improving interpretability while retaining predictive power.

Standard scaling is applied to normalize the features, addressing their wide numerical range (e.g., \(10\)–\(100\ \si{\giga\pascal}\) for shear modulus). Finally, the dataset is split into 160 training and 40 test samples. Correlation analysis is then performed to inspect redundancy among features, aiding subsequent PCA and regularization techniques. This pre-analysis step helps minimize overfitting by identifying and mitigating highly collinear inputs in early trials.

\subsection{Steady-state wear rates}
\noindent For the data analysis, we hypothesize that the wear rate, \(\dot{h}\), remains constant once the system reaches steady-state conditions. To test this hypothesis, DEM simulations are performed for two configurations: (a) varying the plate angle while keeping the particle factory angle fixed at \(45\si{\degree}\), and (b) varying the particle factory angle while keeping the plate angle fixed at \(45\si{\degree}\).~\Cref{fig:simulation_setup} illustrates the geometric configuration of the simulation setup, featuring the plate angle and particle factory angle. In cases (a) and (b), all other parameters like material properties (Young's modulus, Poisson's ratio, coefficient of restitution, static friction, and rolling friction for both particles and equipment), gravitational acceleration, particle size (constant for all particles), initial particle velocity (set to zero), and Archard's wear constant \(k = 10^{-10}\ \si{\per\pascal}\) are kept constant. No relative motion occurred between the plate and the particle factory apart from particle impact and flow due to gravity.
Here, the steady state is defined as the period during which the wear process exhibits a constant rate of change, i.e., particle-plate contacts stabilize following the initial transient phase. In other words, after the initial fluctuations caused by particle deposition and initial contacts, the system evolves into a regime where \(\dot{h}\) is time-invariant.

\begin{figure}[H]
    \centering
        \subfloat[]{\includegraphics[width=0.5\linewidth]{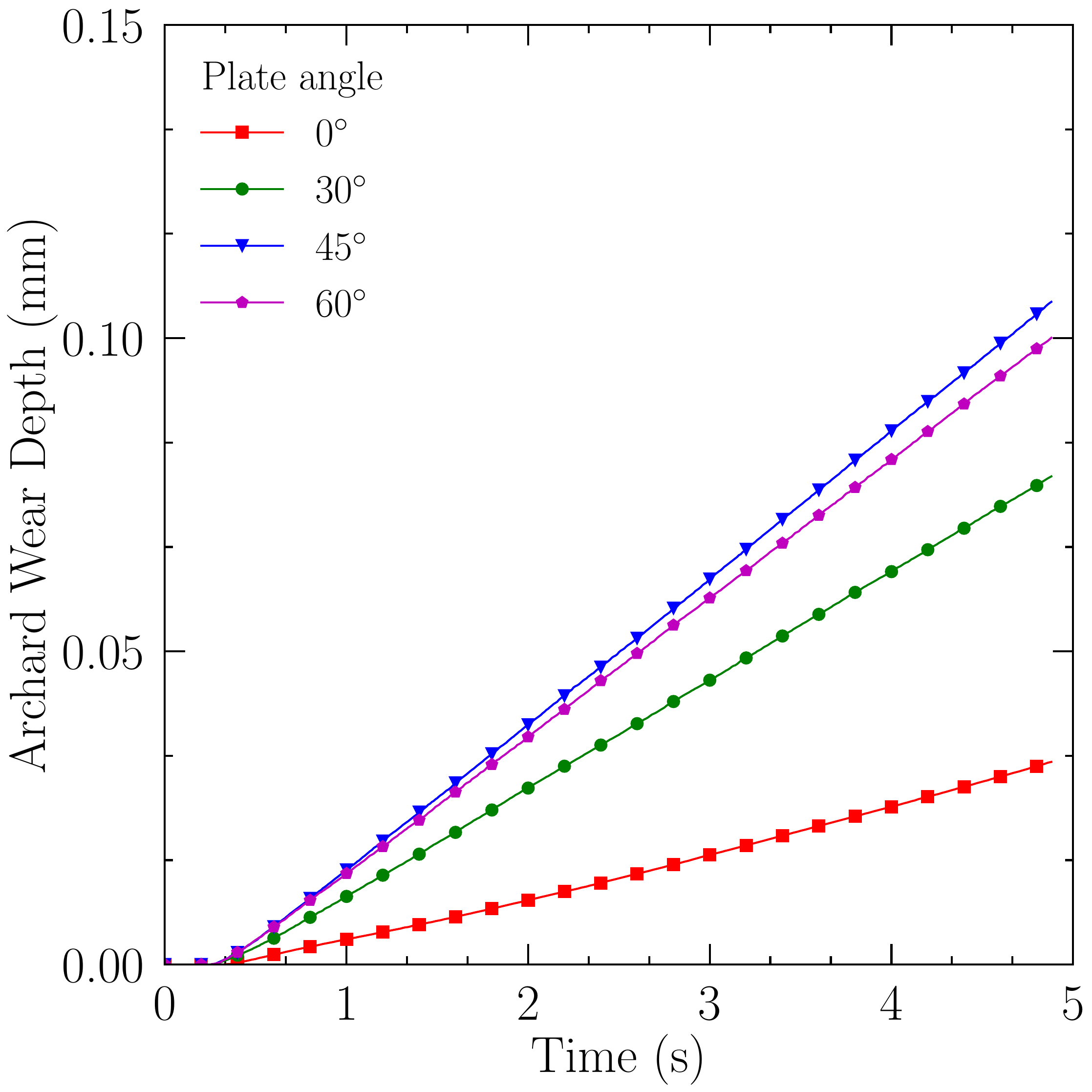}}
        \subfloat[]{\includegraphics[width=0.5\linewidth]{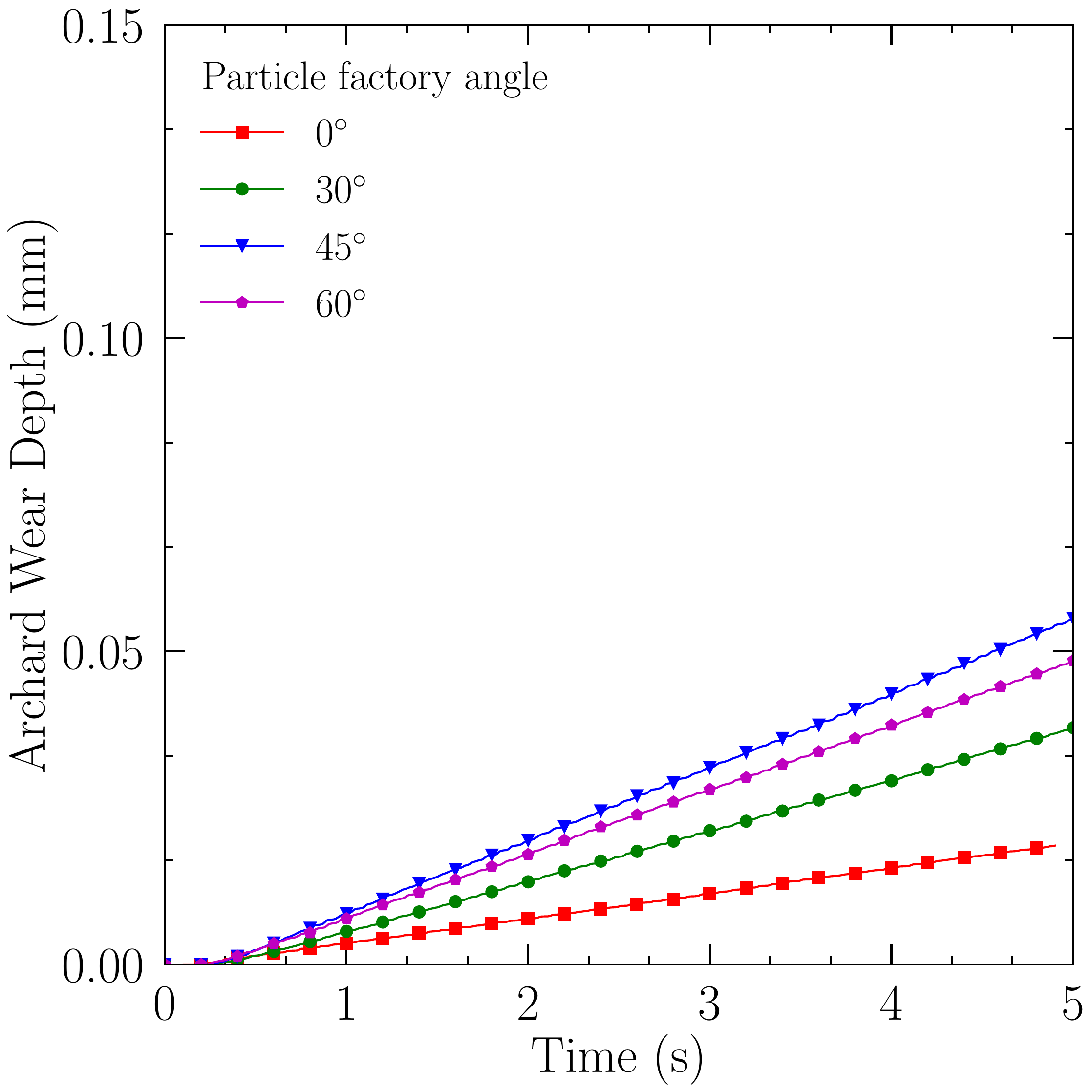}}
    \caption{Archard's wear depth evolution with time (a) varying plate angles (factory angle fixed at 45°) (b) for varying factory angles (plate angle fixed at 45°)}
    \label{fig:wear_rate_consistency}
\end{figure}

\Cref{fig:wear_rate_consistency} compares Archard's wear depth with time. The wear depth increased linearly with time across all simulation runs with \(R^2\) values exceeding 0.98, indicating an excellent fit to the linear model~\citep{jayasundara2022predicting}. This underpins the assumption that the wear rate, \(\dot{h}\), is constant under steady-state conditions. Notably, the wear rate is linear; the variation in steady-state wear rate for geometric parameters, like plate angle or particle factory angle, is not strictly linear. This observation suggests that purely linear models may be limited in accuracy in capturing the influence of these geometrical parameters, thereby motivating the need to explore nonlinear or interaction-based modeling approaches. 

\section{Results and discussion}\label{resultsAndDiscussions}
\subsection{Linear regression: Feature insights and limitations}
\begin{figure}[H]
    \centering
    \subfloat[]{\includegraphics[width=0.6\linewidth]{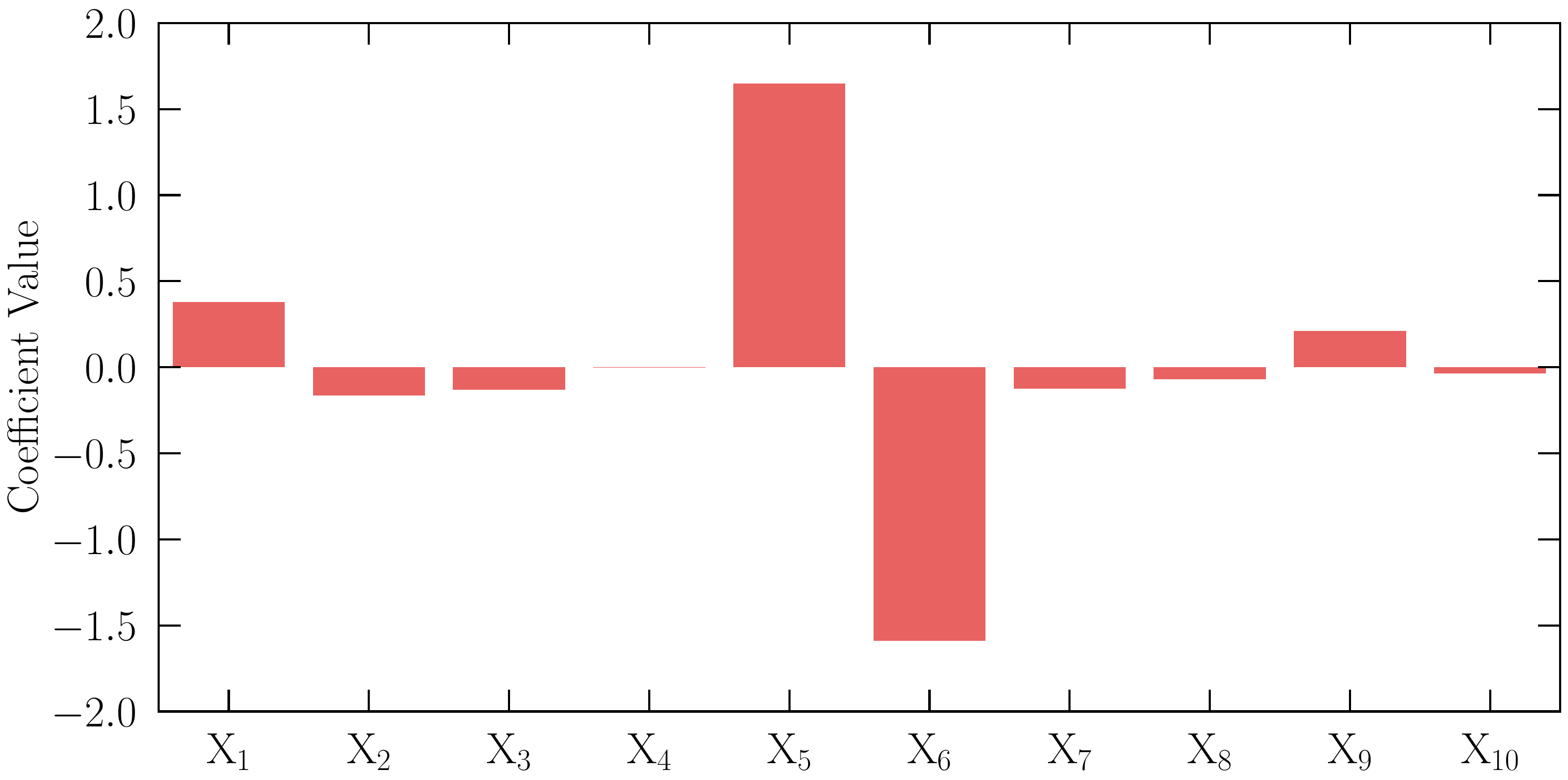}}\\
    \subfloat[]{\includegraphics[width=0.6\linewidth]{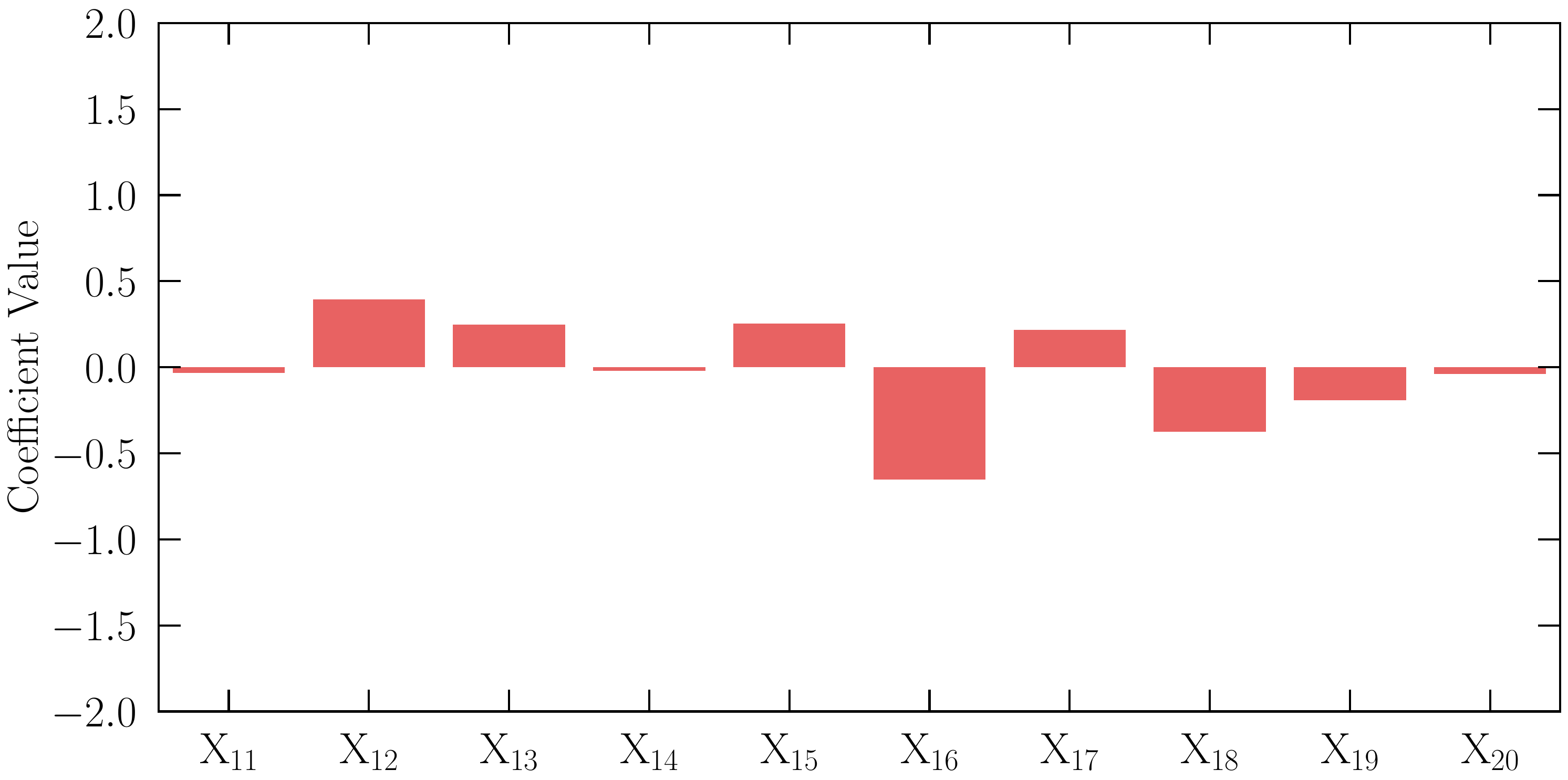}}
    \caption{Bar plots of linear regression coefficients showing relative importance of input features (a) \(X_1-X_{10}\) (b) \(X_{11}-X_{20}\).}
    \label{fig:regression_bar_plots_combined}
\end{figure}

\noindent Linear regression is initially employed to relate input features to the wear rate.~\Cref{tab:feature_mapping} delineates the physical meaning of each feature in the model. The model achieved an \( R^2 \) score of 0.68, indicating moderate predictive accuracy. Nonetheless, it provided valuable insights into the relative importance of the input parameters.~\Cref{fig:regression_bar_plots_combined} and~\Cref{eq:linear_regression} presents the regression coefficients for the significant variables, including the Young's modulus of particle \((X_1)\), Young's modulus of equipment \((X_5)\), Archard's wear constant \((X_{12})\), and particle's \(Y\) velocity \((X_{16})\). These four features have the highest positive weights, signifying a strong influence on wear rate behavior.

The Young's modulus of the equipment \((X_5)\) has a positive effect, as a higher modulus increases the contact normal force \(F^\text{n}\), which in turn increases wear due to higher contact stresses. In contrast, the shear modulus \((X_6)\) exhibits a negative coefficient, as an increase in shear modulus increases tangential stiffness, which reduces sliding distance and consequently results in less wear. This difference implies that, although moduli relate to stiffness, they influence impact interactions in distinct ways. The Young's modulus of the particle \((X_1)\) showed a moderate positive weight, indicating that harder particles generate greater contact forces, increasing wear.

The coefficient of restitution \((X_9)\) exhibited a reasonable positive influence, likely because more energetic rebounds increase surface interactions. Similarly, the plate angle \((X_{13})\) also had a positive contribution, suggesting that steeper angles enhance the normal force components, resulting in increased wear. Some features, such as the effective Young's modulus \((X_{17})\) and effective shear modulus \((X_{18})\), exhibited negative coefficients. This may indicate redundancy, as they strongly correlate with the individual elastic parameters already included in the model. The regression model is as follows:
\begin{equation}
\begin{aligned}
\text{Wear Rate} =\ & -0.019 + 0.380X_1 + 0.164X_2 - 0.130X_3 + 0.001X_4 + 1.64X_5 \\
& - 1.59X_6 - 0.123X_7 - 0.071X_8 + 0.29X_9 - 0.035X_{10} - 0.034X_{11} \\
& + 0.394X_{12} + 0.246X_{13} - 0.020X_{14} + 0.255X_{15} - 0.653X_{16} \\
& - 0.218X_{17} - 0.376X_{18} - 0.193X_{19} - 0.039X_{20}
\end{aligned}
\label{eq:linear_regression}
\end{equation}

While the linear model provided interpretable weights, it also has limitations. Multicollinearity is evident, especially between features like Young's modulus and the effective modulus, which are theoretically dependent. Additionally, the damping ratio \((X_{19})\) and the damping coefficient \((X_{20})\) showed low weights, suggesting minimal independent influence or overlap with other dynamic parameters. These redundancies highlight the need for dimensionality reduction and feature selection to achieve more robust modeling.

\begin{figure}[H]
    \centering
    \includegraphics[width=0.6\textwidth]{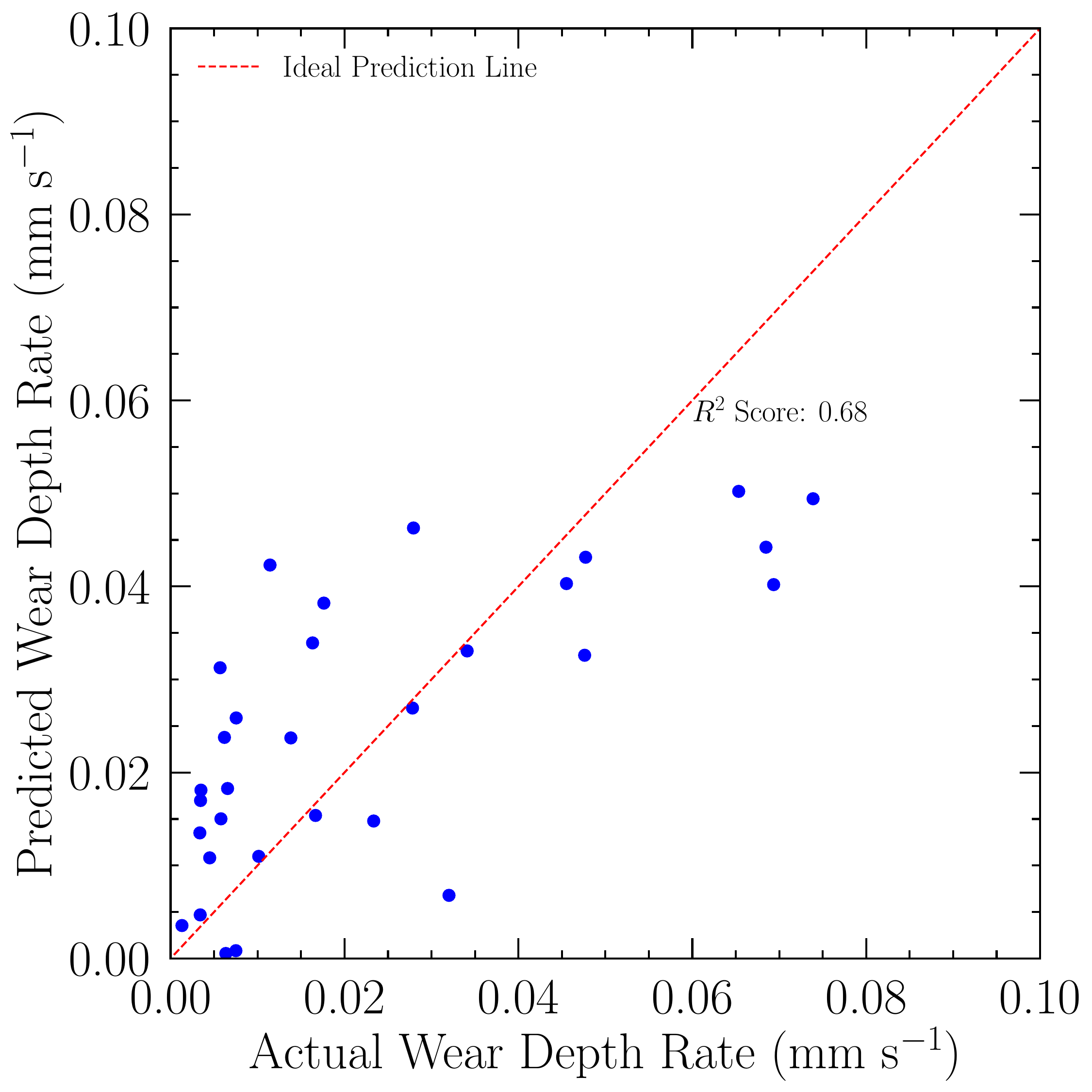}
    \caption{Comparison of predicted versus actual wear depth rates \((\si{\milli\meter\per\second})\) using the developed model. The red dashed line represents the ideal prediction line.}
    \label{fig:linreg_residuals}
\end{figure}

\Cref{fig:linreg_residuals} compares the predicted wear depth rate with the actual wear depth rate. It reveals a non-constant variance and indicates heteroscedasticity. This finding further diminished the reliability of the linear model's predictions, leading to the adoption of regularization techniques and nonlinear models in subsequent stages.

\subsection{Feature selection via regularization} \label{subsec:regularization}
\noindent Two regularization techniques\textendash Lasso (L1) and Ridge (L2)\textemdash are applied to improve model interpretability and manage multicollinearity. Both methods simplified the model while preserving predictive performance, achieving an \( R^2 \) score of approximately 0.68.

\begin{figure}[H]
    \centering
    \subfloat[]{\includegraphics[width=0.6\linewidth]{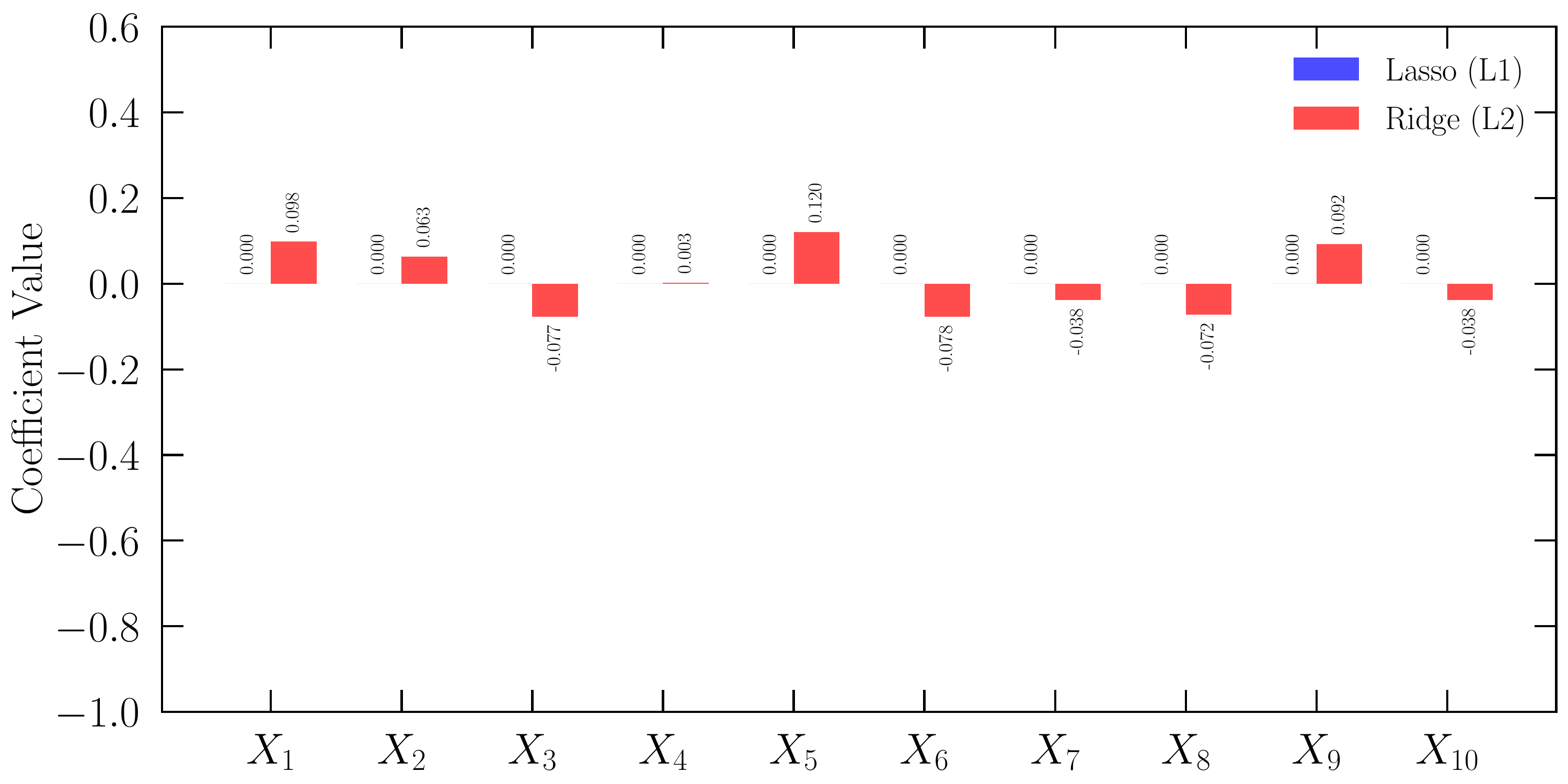}}\\
    \subfloat[]{\includegraphics[width=0.6\linewidth]{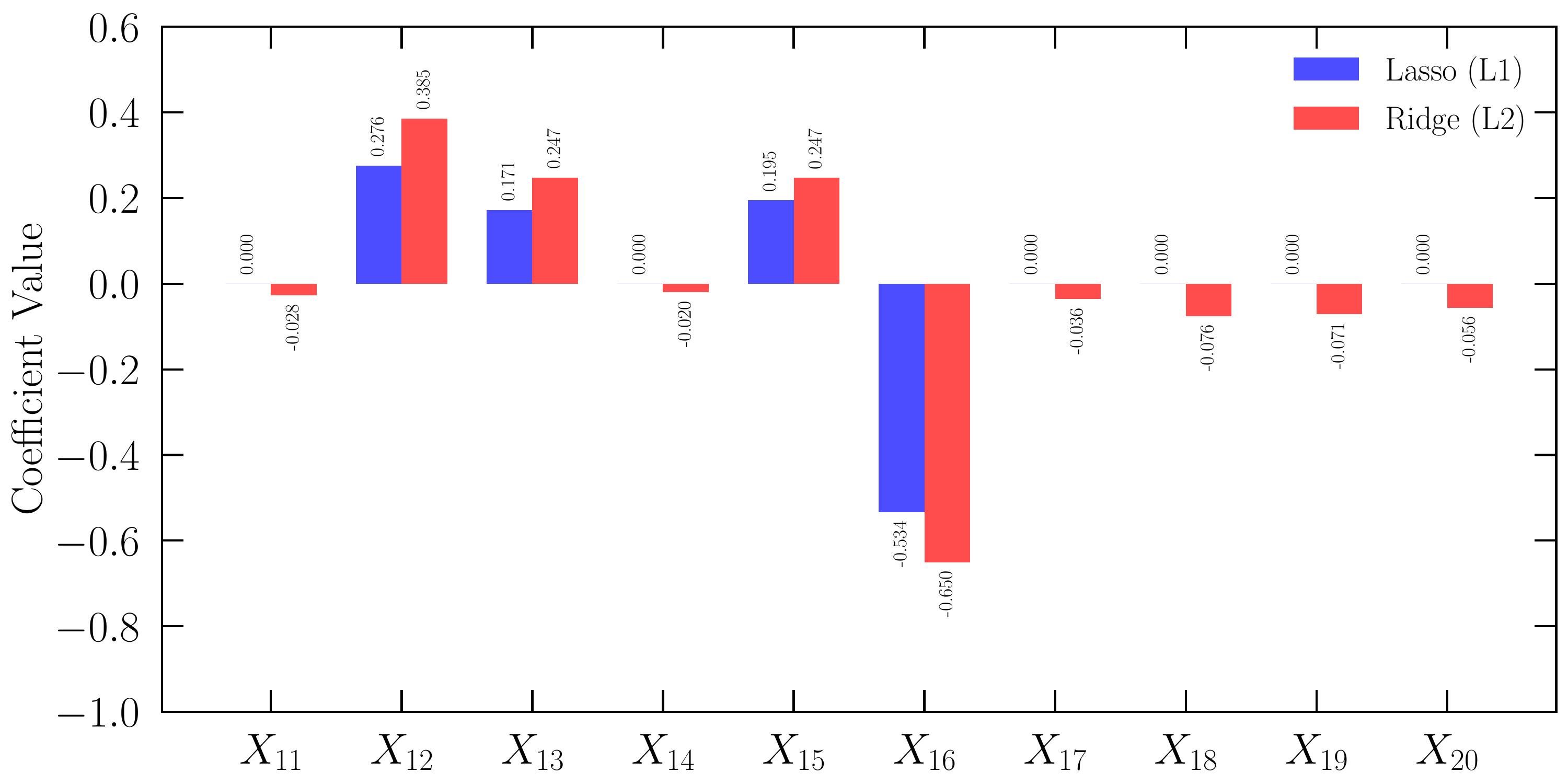}}    
    \caption{Comparison of Ridge and Lasso coefficients for standardized features. A visual representation of weight strength (a) \(X_1-X_{10}\) (b) \(X_{11}-X_{20}\).}
    \label{fig:regularization_comparison}
\end{figure}

\Cref{fig:regularization_comparison} illustrates the coefficients of standardized features for Ridge and Lasso regularization techniques. Lasso regression reduces the coefficients of less important features to zero, effectively selecting only the most relevant predictors. In this study, the Lasso model retained four key features: Archard's wear constant \((X_{12})\), plate angle \((X_{13})\), particle size \((X_{15})\), and particle's \(Y\) velocity \((X_{16})\). The regression model obtained from Lasso is presented below:
\begin{equation}
    \text{Wear Rate}_{\text{L1}} = 0.276X_{12} + 0.171X_{13} + 0.195X_{15} - 0.534X_{16} + \sum \text{other terms}
\end{equation}

Each of the retained features has a clear physical interpretation. Archard's wear constant is a direct scaling factor in the classical wear model, which explains its strong positive coefficient. An increased plate angle enhances the sliding velocity, leading to greater wear. Larger particle sizes are linked to higher contact stresses, and a more negative particle's \(Y\) velocity (indicating a higher impinging velocity) results in more aggressive wear.

Ridge regression, in contrast, retained all the original features from linear regression while reducing the magnitude of less influential coefficients. This approach is valuable when it is important to maintain all physical parameters while minimizing the risk of overfitting. The Ridge model also highlighted the same dominant features as Lasso\textemdash\(X_{12}\), \(X_{13}\), \(X_{15}\), and \(X_{16}\) but with higher coefficients. The core part of the Ridge regression equation is presented below:
\begin{equation}
    \text{Wear Rate}_{\text{L2}} = 0.385X_{12} + 0.247X_{13} + 0.247X_{15} - 0.650X_{16} + \sum \text{other terms}
\end{equation}

Despite their distinct approaches to handling non-essential variables, Lasso and Ridge regression consistently identified the same four dominant predictors. This convergence indicates a strong physical and statistical relevance of these parameters. Lasso is particularly beneficial when model simplicity is preferred, whereas Ridge is advantageous when maintaining feature completeness is essential.

\subsection{Principal component analysis for feature grouping}
\label{subsec:pca}
\noindent Further analysis of the relationships among features and validation of the results from regularization are conducted using Principal Component Analysis (PCA). This method transforms correlated input features into a set of linearly uncorrelated components, known as principal components (PCs), which capture the maximum variance in the data.

\begin{figure}[H]
    \centering
    \subfloat[]{\includegraphics[width=1.0\linewidth]{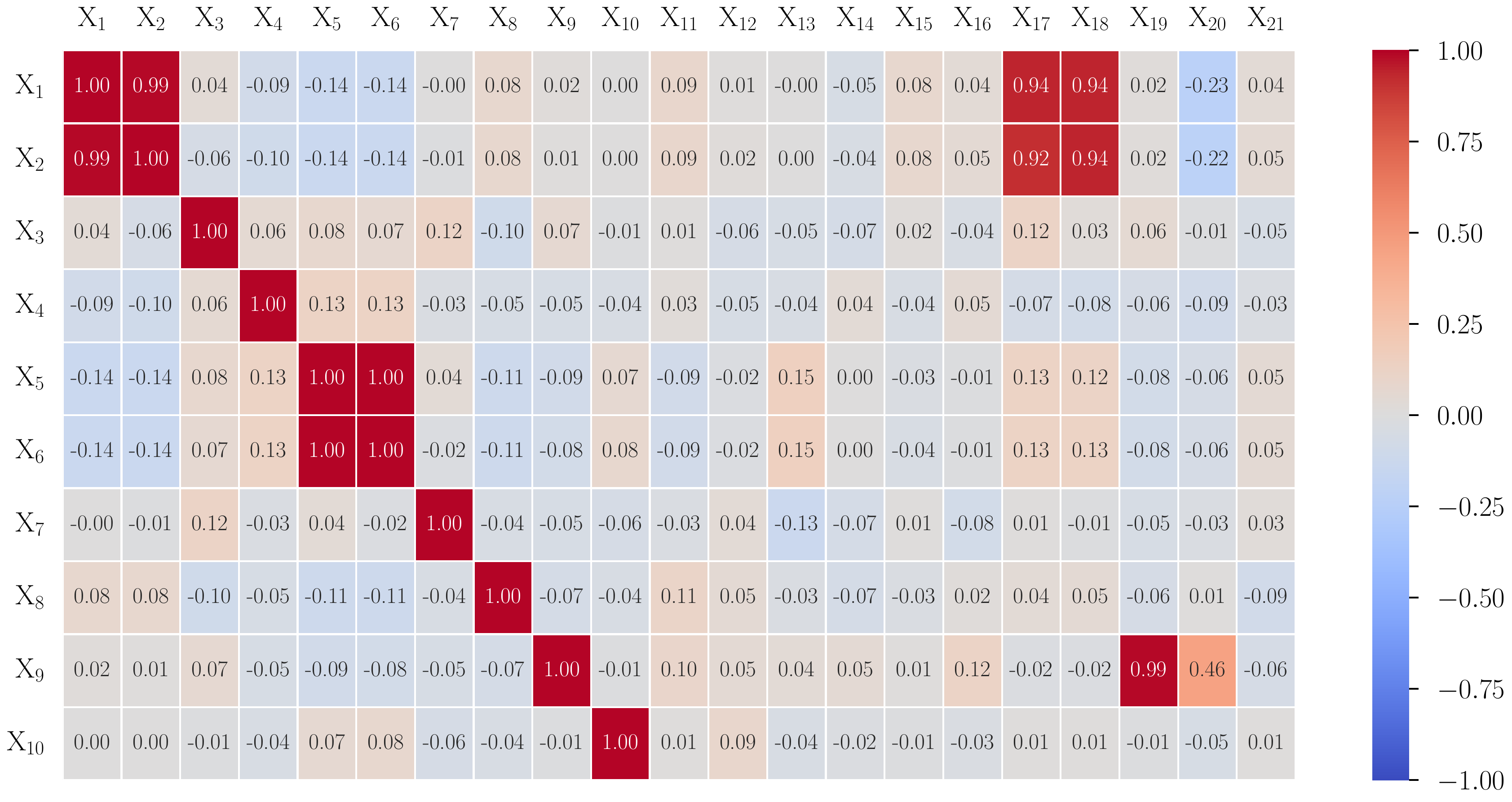}}\\
    \subfloat[]{\includegraphics[width=1.0\linewidth]{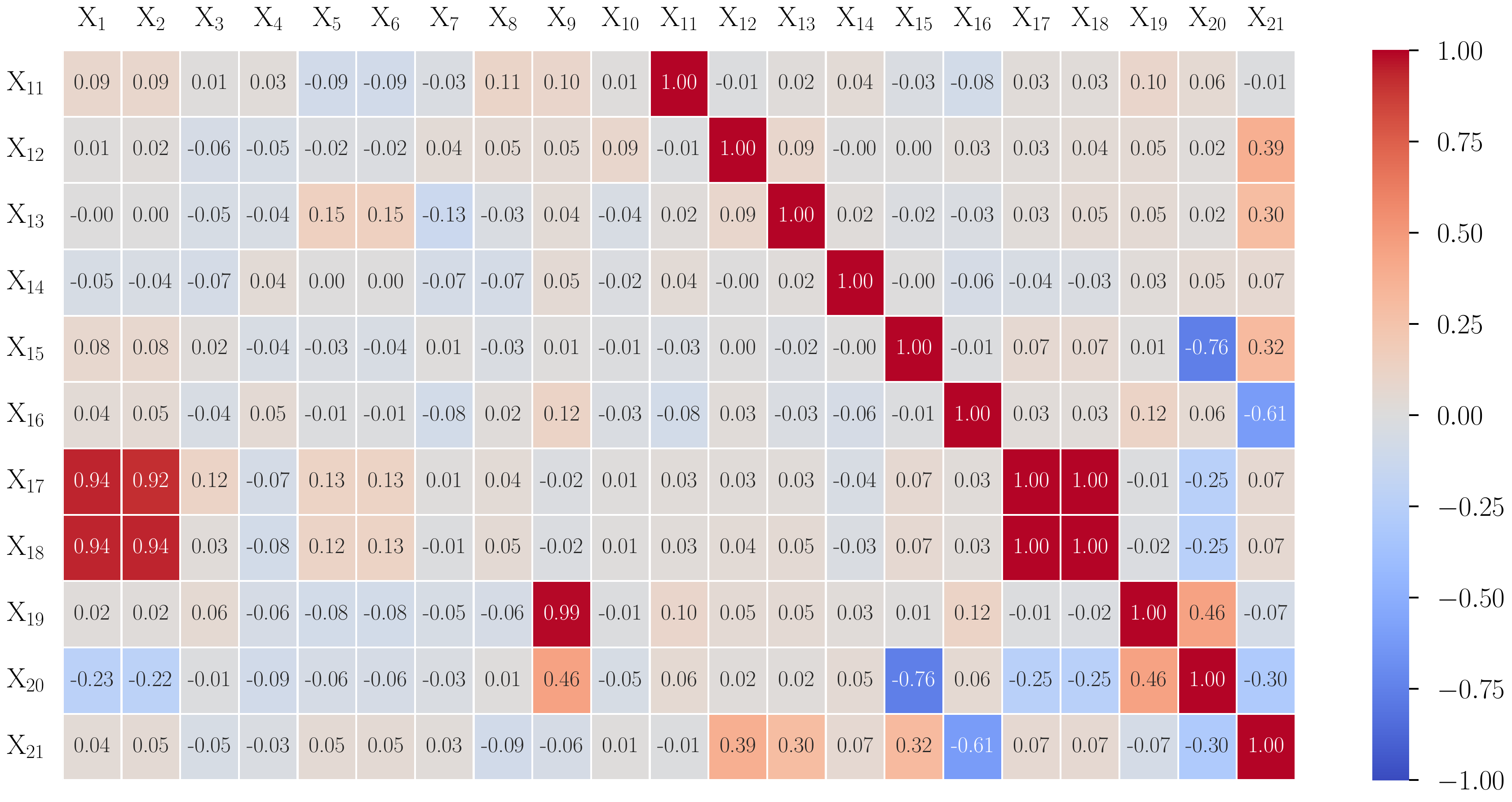}}
    \caption{Correlation heatmap for feature selection using principal component analysis (a) \(X_1-X_{10}\) (material and equipment properties) (b)Correlation heatmap for features \(X_{11}-X_{20}\), representing interaction and geometric parameters. The variable \(X_{21}\) denotes the wear rate of the plate (i.e., target variable).}
    \label{fig:corr_heatmap1}
\end{figure}

\begin{figure}[H]
    \centering
    \subfloat[]{\includegraphics[width=1.0\linewidth]{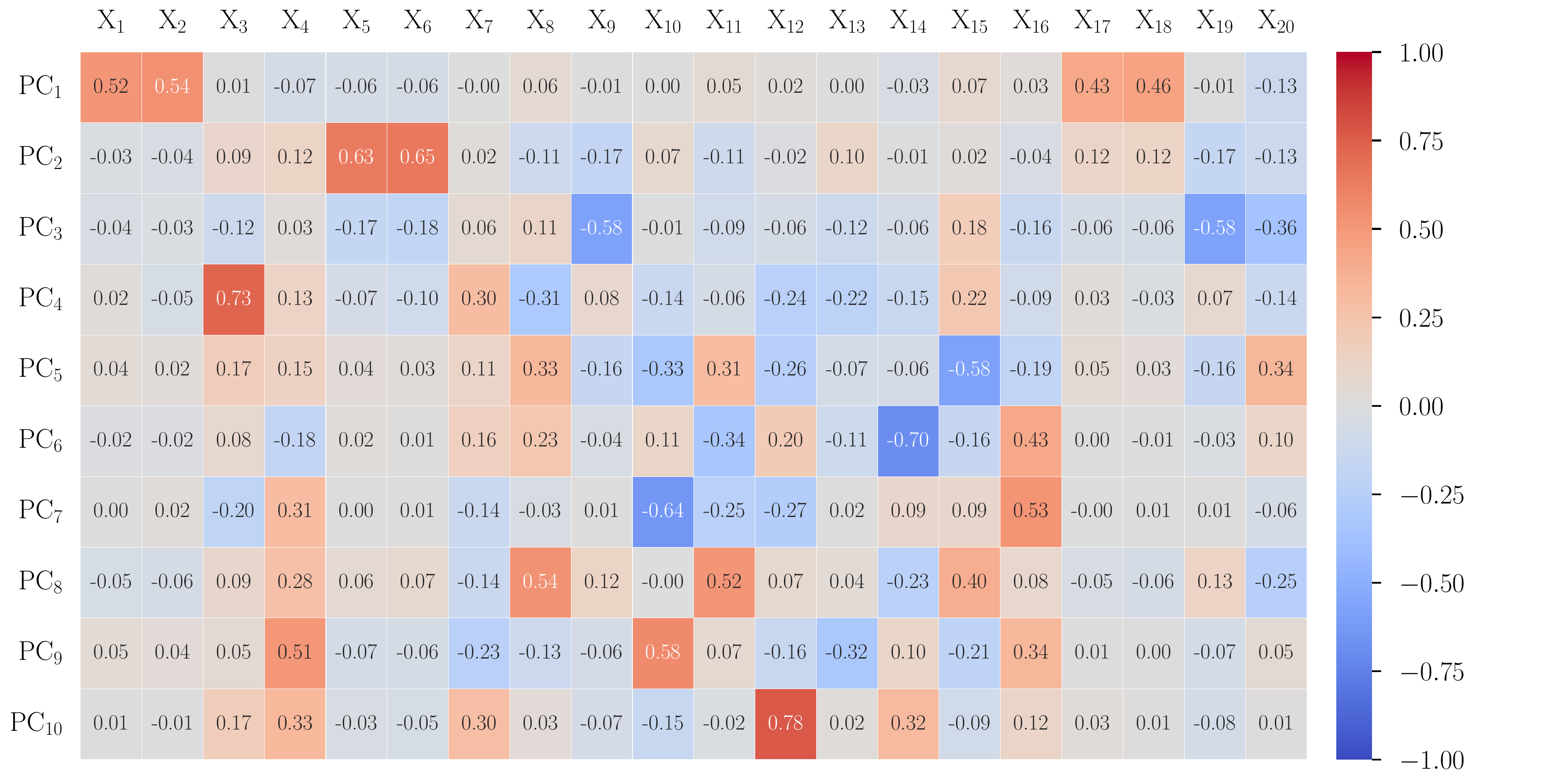}}\\
    \subfloat[]{\includegraphics[width=1.0\linewidth]{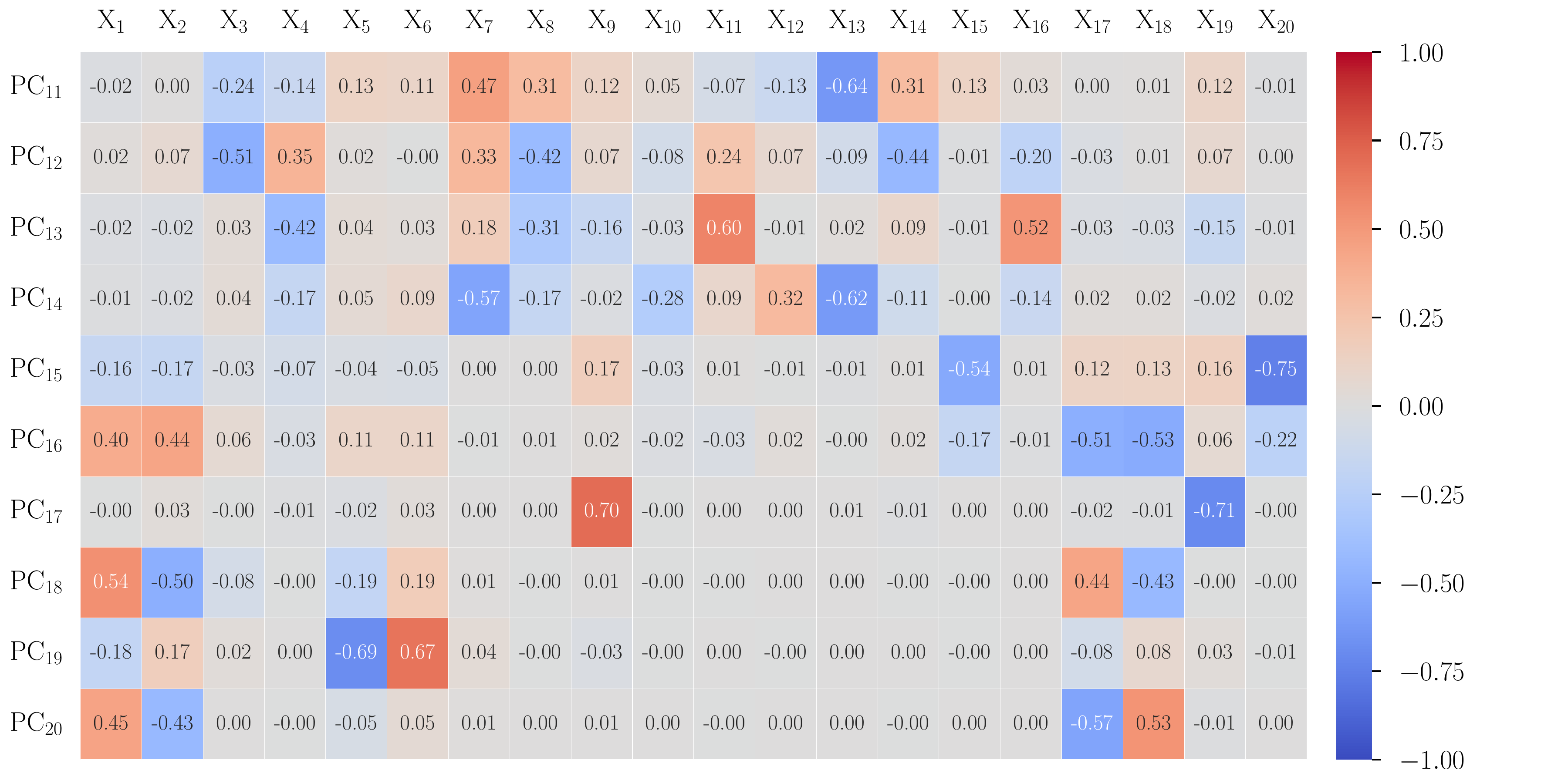}}
    \caption{Heatmap of principal component loadings for variables \(X_1-X_{20}\). The color intensity represents the strength and direction of each variable’s contribution to the (a) \(PC_1-PC_{10}\) (b) \(PC_{11}-PC_{20}\).  The color red indicates positive loadings, and blue indicates negative loadings.}
    \label{fig:pca_loadings1}
\end{figure}

\Cref{fig:corr_heatmap1} shows the correlation heatmaps for features \(X_1-X_{10}\) (material and equipment properties) and \(X_{11}-X_{20}\) (interaction and geometric properties). The variable \(X_{21}\) denotes the wear rate of the plate (target variable). High correlation between features—particularly stiffness parameters—indicates redundancy in the dataset, reinforcing the need for dimensionality reduction or selection. PCA revealed meaningful groupings among the features. For instance, \(PC_1\) is dominated by modulus-related parameters of Particle \(X_1, X_2, X_{17}, X_{18}\) and \(PC_2\) is dominated by modulus-related parameter of Plate \(X_5, X_6\), see~\Cref{fig:pca_loadings1}. The coherence between PCA and regularization reinstates that the stiffness parameters behave similarly and may be grouped. In contrast, the kinematic and wear-related parameters operate independently. In particular, PCA further supported the selection of Young's modulus over shear modulus based on consistent variance contribution.

\begin{figure}[H]
    \centering
    \includegraphics[width=0.6\linewidth]{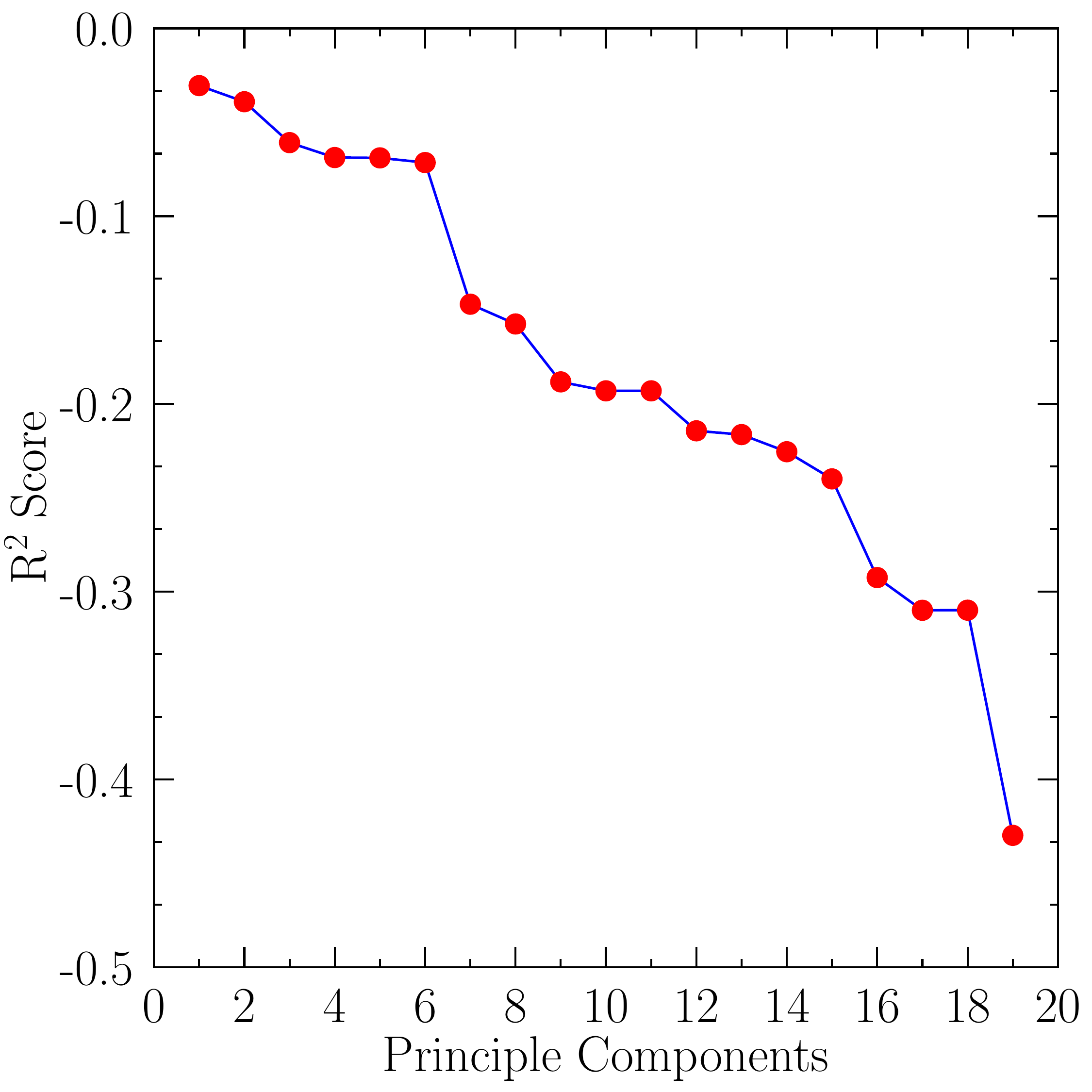}
    \caption{Comparison of the residual \(R^2\) values for linear regression using increasing principal components.}
    \label{fig:pca_regression}
\end{figure}

\noindent \textbf{Remark:} The insights from PCA for feature groupings seemed promising, but their application for regression analysis led to poor results. A linear regression model trained on PCA-transformed features with an \(R^2\) of -0.21 underperforms baseline models, see~\Cref{fig:pca_regression}. The poor performance is attributed to the transformation, which alters physically meaningful relationships among features, and linear combinations of PCs cannot capture the nonlinear nature of wear mechanisms. Consequently, PCA transformation is not used for model input. Instead, PCA is only applied to support and validate feature selection by identifying groups of related variables.

\subsection{Feature selection}
\noindent The Combined results of regularization and PCA reveal a final set of twelve features selected for further modeling. These features are identified as important based on high regression weights, strong PCA loadings, or both. The selected features are:
\begin{enumerate}[labelindent=1em, labelsep=0.25cm, leftmargin=*]
    \item Young's Modulus (particle) \((X_1)\)
    \item Shear Modulus (particle) \(X_2\)
    \item Poisson's Ratio (particle) \((X_3)\)
    \item Young's Modulus (Equipment) \((X_5)\)
    \item Shear Modulus (Equipment) \(X_6\)
    \item Poisson's Ratio (Equipment) \((X_7)\)
    \item Coefficient of Restitution (Interaction) \((X_9)\)
    \item Archard's Wear Constant \((X_{12})\)
    \item Plate Angle \((X_{13})\)
    \item Particle Size \((X_{15})\)
    \item Particle \(Y\) Velocity \((X_{16})\)
    \item Damping Coefficient \((X_{20})\)
\end{enumerate}

The Archard's wear constant, plate angle, particle size, and particle velocity strongly influenced the wear rate. The remaining eight are retained for their moderate but consistent contributions across regularization and PCA, and their physical relevance. This selected set provides a robust and interpretable input for the machine learning models used in the following analysis stage.

\subsection{Decision tree performance analysis}
\begin{figure}[H]
    \centering
    \includegraphics[width=1\textwidth]{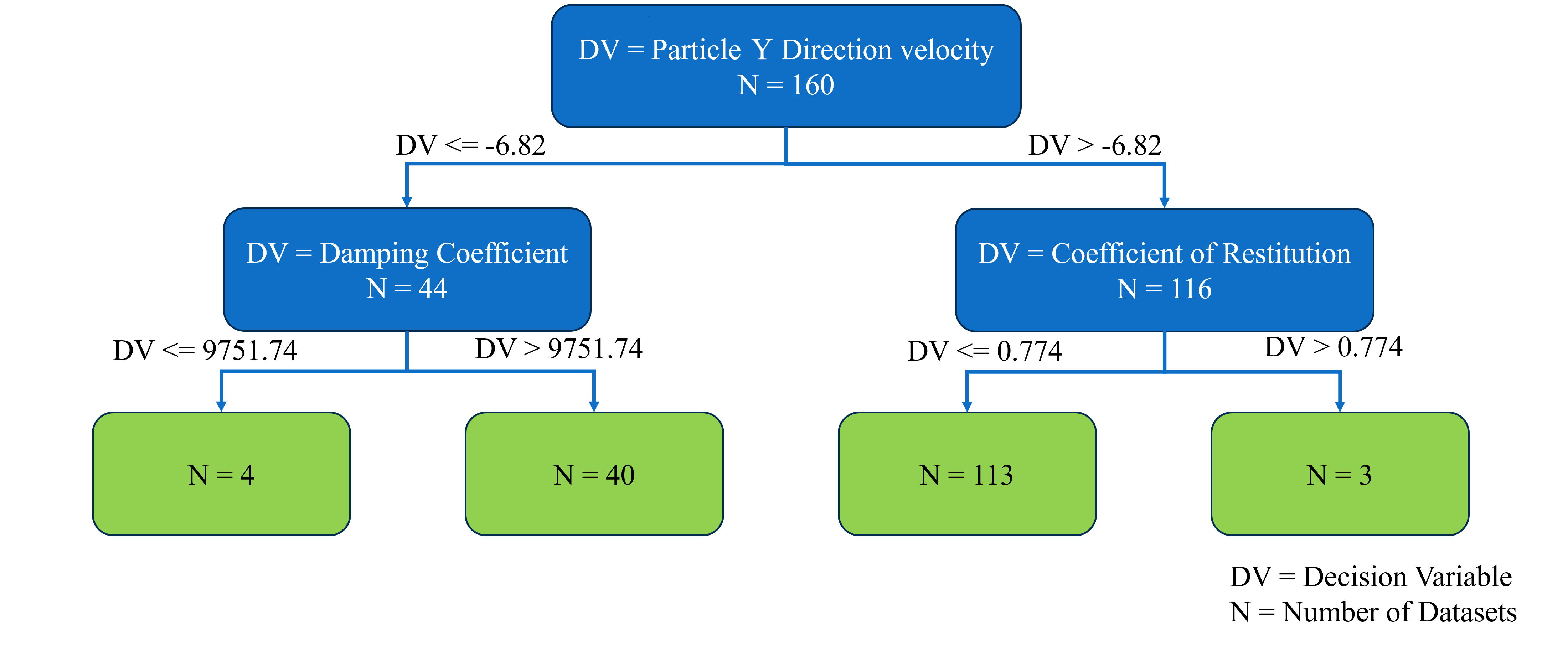}
    \caption{Conceptual illustration of a two-level decision tree, where DV is the decision variable. It demonstrates how the splits occur in a decision tree based on optimal feature thresholds.}
    \label{fig:decision_tree_illustration}
\end{figure}

\noindent A decision tree regressor is implemented and systematically analyzed to explore the ability of tree-based models to capture nonlinear relationships in the data.~\Cref{fig:decision_tree_illustration} provides a conceptual illustration of a two-level decision tree, demonstrating how this model operates by sequentially splitting the data based on optimal feature thresholds. This visual representation helps clarify the fundamental mechanism by which decision trees partition the feature space and make predictions.

\begin{figure}[H]
    \centering
    \includegraphics[width=0.6\textwidth]{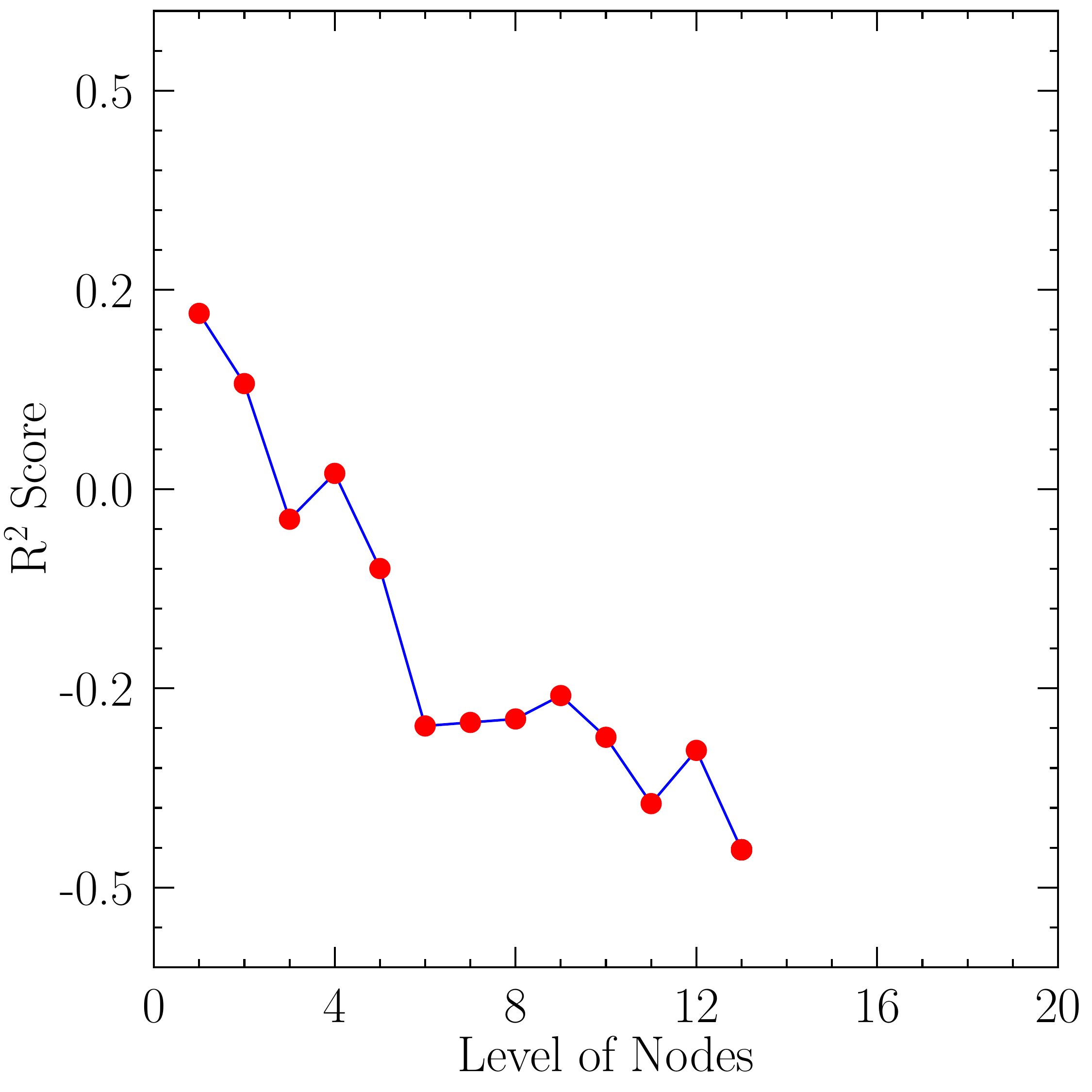}
    \caption{ Comparison of the \(R^2\) score versus decision tree depth, highlighting the model performance with complexity (level of nodes).}
    \label{fig:r2_vs_tree_depth}
\end{figure}

The impact of model complexity on predictive accuracy is quantitatively assessed by varying the tree depth and recording the corresponding \(R^2\) scores on the test set.~\Cref{fig:r2_vs_tree_depth} illustrates the best performance achieved with very shallow trees (depth 1 or 2), yielding \(R^2\) values around 0.2. This indicates that the simplest decision trees capture only the most basic trends in the data.

As tree depth increased beyond 4 or 5, the \(R^2\) score dropped sharply, reaching negative values (approximately \(-0.25\) at depth 7). This decline indicates overfitting, where the model becomes too complex, memorizing the training data but failing to generalize to new samples. Although there is a slight uptick in \(R^2\) around depth 12, scores remain negative for all depths greater than 4, confirming that increased complexity consistently worsens predictive performance.

The decision tree regressor showed limited effectiveness for this wear prediction task. While very shallow trees offered modest predictive power, deeper trees experienced severe overfitting, resulting in worse predictions than a simple average. The model's inability to generalize, evidenced by negative \(R^2\) values at most depths, makes it the least practical approach among those evaluated. These findings highlight the need to transition to more advanced nonlinear models, such as Artificial Neural Networks (ANNs), which are better suited for capturing the complex relationships in the dataset.

\subsection{GA-optimized ANN}
\begin{figure}[H]
    \centering
    \includegraphics[width=0.8\textwidth]{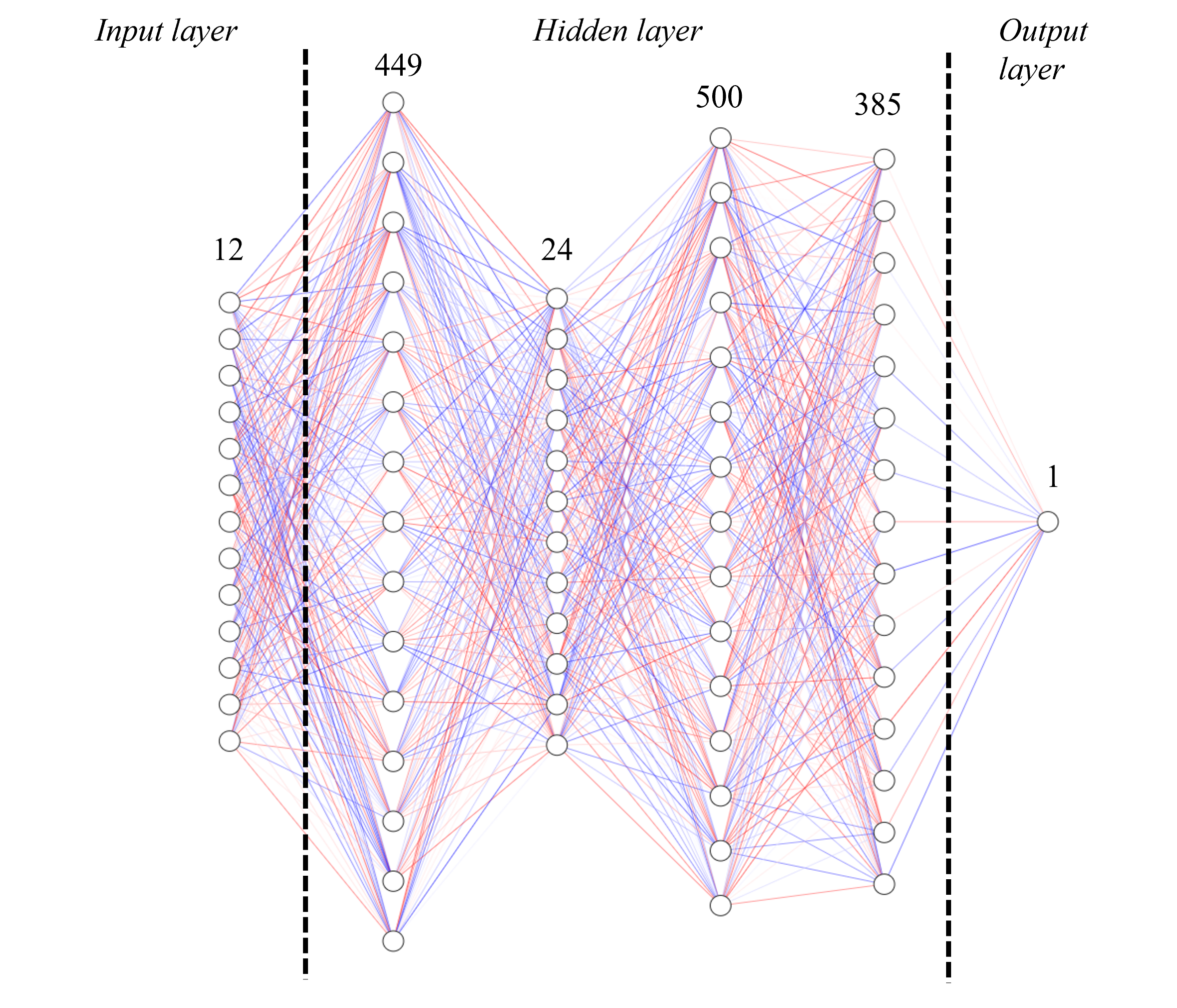}
    \caption{Schematic representation of the optimized neural network architecture with four hidden layers determined by a genetic algorithm.}
    \label{fig:ga_ann_architecture}
\end{figure}

\noindent The integration of a GA to optimize the architecture of the ANN led to a substantial improvement in predictive accuracy for wear rate estimation. The GA identified an unconventional yet highly effective network structure by systematically exploring variations in the number of hidden layers and neurons per layer.~\Cref{fig:ga_ann_architecture} shows a schematic of the optimized neural network architecture with four hidden layers determined by a genetic algorithm. In this study, we used an architecture comprised of four hidden layers with neuron counts of 449, 24, 500, and 385, respectively. This non-monotonic configuration diverges from traditional ANN designs, often employing a constant or gradually decreasing number of neurons across layers. It highlights the capacity of evolutionary algorithms to uncover novel solutions that may not be intuitive to human designers.

The model is trained and evaluated using a dataset split of 70\% and 30\% for training and testing, respectively, ensuring robustness in its generalization capability. All hidden layers utilized the Rectified Linear Unit (ReLU) activation function, contributing to efficient learning and mitigating the vanishing gradient problem. The GA optimization process is executed over 25 generations to maximize the coefficient of determination ($R^2$) on the validation set.

The optimized GA-ANN model achieved an \(R^2\) score of 0.91 on the test dataset, indicating that it could explain approximately 91\% of the variation in wear rates based on the input features. The scatter plot comparing actual and predicted wear rates demonstrated that most predictions followed the ideal line, confirming the model's high predictive power across the full range of observed wear values. The balanced distribution of points around the identity line further suggested the absence of systematic overestimation or underestimation.
\begin{figure}[H]
    \centering
    \includegraphics[width=0.6\textwidth]{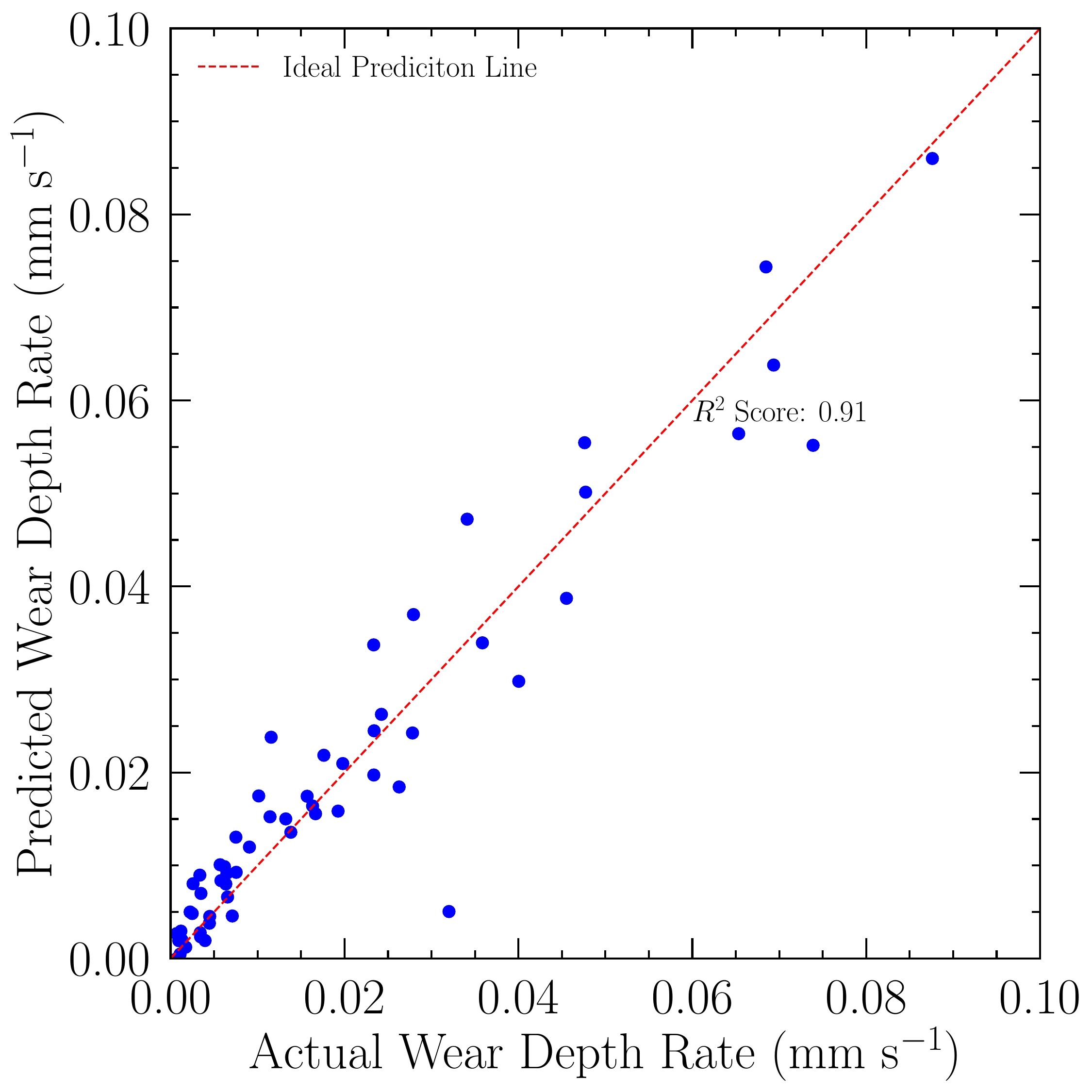}
    \caption{Comparison of actual versus predicted wear rates using the GA-optimized ANN model on the test dataset.}
    \label{fig:actual_vs_predicted}
\end{figure}

\begin{figure}[H]
    \centering
    \includegraphics[width=0.75\textwidth]{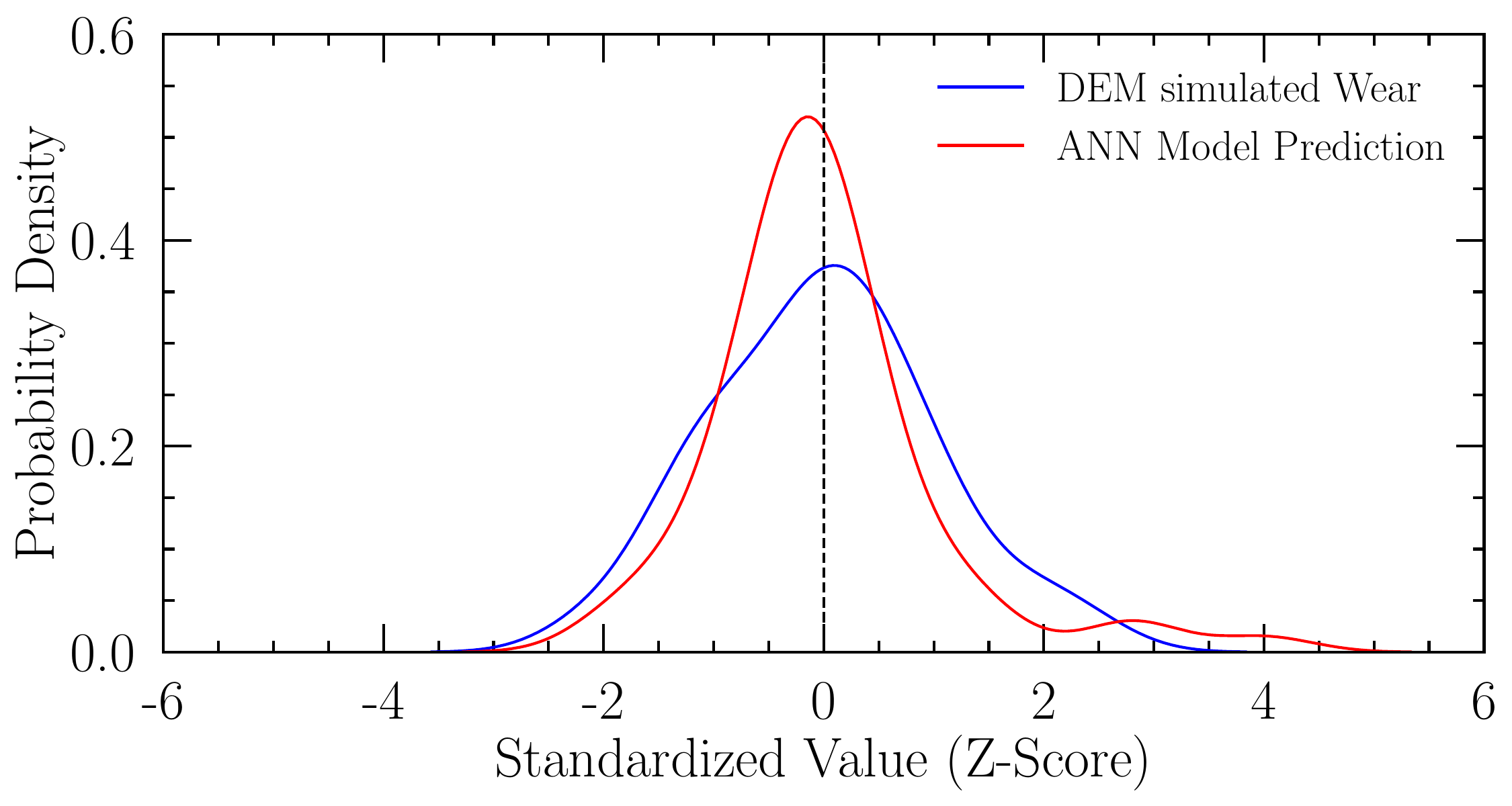}
    \caption{Statistical distribution of standardized values (Z-score) for wear and ANN model predictions.}
    \label{fig:distribution_comparison}
\end{figure}

\Cref{fig:distribution_comparison} illustrates the statistical distributions of the wear data from the DEM simulations and the predictions from the ANN model. The analysis reveals a significant reduction in variance and a pronounced peak near zero in the wear distribution. Furthermore, \Cref{fig:distribution_comparison} demonstrates that most prediction errors are smaller and symmetrically distributed, with only a few outliers present. This probability distribution reinforces the model's accuracy and indicates its unbiased nature in predictions.

The GA-ANN model significantly outperformed multiple linear regression, \( R^2 \) of 0.68, and the decision tree regressor, which achieved a maximum \( R^2 \approx 0.58\) compared to other methods. The superior performance of the GA-ANN is attributed to the combination of careful feature selection, the capacity of genetic algorithms to optimize complex network structures, and the inherent ability of deep learning models to capture intricate, nonlinear relationships within the data. These results underscore the value of evolutionary optimization in neural network design for predictive maintenance applications in bulk material handling equipment.

\section{Summary and conclusions}\label{summaryAndConclusion}
\noindent This study presents a machine learning-based framework for predicting particle-induced Archard's wear using DEM simulation data. Integrating physics-informed simulation data with data-driven machine learning techniques provides a powerful tool for accurate wear prediction. This framework reduces dependence on extensive experimental testing and opens pathways for real-time wear monitoring and design optimization in tribological systems. The major conclusions are delineated below.
\begin{itemize}[labelindent=1em, labelsep=0.5cm, leftmargin=*]
    \item A few machine learning techniques are evaluated\textemdash multiple linear regression and decision tree regressors. The GA-ANN demonstrated better performance, with an \(R^2\) score of 0.91 on the test dataset. The results underpin ANN to establish complex, nonlinear relationships between wear and parameters like material properties, process conditions, and geometric features.

    \item The ANN model showed consistent and balanced predictions across the full range of wear values, with low residual errors symmetrically distributed around zero. This confirms the robustness and generalization capability of ANN. The performance improvement is attributed to effective feature selection, a genetic algorithm for hyperparameter tuning, and the intrinsic ability of deep learning methods to capture complex patterns within the dataset.    
\end{itemize}

Future work will extend this methodology to accommodate more complex geometries, like curved plates and configurations featuring multiple connected linear plates. Additionally, since DEM simulations inherently represent surfaces as assemblies of planar facets, these can be effectively approximated as interconnected plates. Incorporating facet-based geometries into the current framework will allow for a more generalized and scalable framework for wear prediction, broadening the model's applicability across various surface types and operating conditions. This progression promises to refine the insights gained from the simulations and improve predictive capabilities in practical scenarios.

\nolinenumbers
\section*{Acknowledgements}
The authors gratefully acknowledge Altair Engineering India for providing access to the EDEM licenses. We also sincerely thank Mr. Prasad Avilala from Altair Engineering India for his valuable technical support. Chandan P. further acknowledges Altair Engineering India for offering an internship opportunity that contributed to this work.

\section*{Data Availability}
The data and code used in this study are publicly available. The code is available on a \href{https://github.com/ChandanPrasanna/Machine-Learning-for-determining-partical-based-wear}{GitHub repository}. The repository includes all relevant scripts and documentation necessary for reproducing the results presented in this study.

\section*{Conflict of Interest}
The authors declare that they have no conflict of interest.

\newpage
\setcounter{section}{0}
\setcounter{page}{1}
\setcounter{figure}{0}
\setcounter{equation}{0}
\renewcommand{\thesection}{S\arabic{section}}
\renewcommand{\thepage}{S\arabic{page}}
\renewcommand{\thetable}{S\arabic{section}.\arabic{table}}
\renewcommand{\thefigure}{S\arabic{section}.\arabic{figure}}
\renewcommand{\theequation}{S\arabic{section}.\arabic{equation}}

\setcounter{affn}{0}
\resetTitleCounters

\makeatletter
\let\@title\@empty
\makeatother

\title{Supplementary Material\\ A DEM-driven machine learning framework for abrasive wear prediction}

\makeatletter
\renewenvironment{abstract}{\global\setbox\absbox=\vbox\bgroup
  \hsize=\textwidth\def\baselinestretch{1}%
  \noindent\unskip\textbf{Contents}
 \par\medskip\noindent\unskip}
 {\egroup}


\startlist{toc}
\begin{abstract}
\vspace{-48pt}
\printlist{toc}{}{\section*{}}
\end{abstract}
\maketitle
\section*{}
\parindent0pt

\setcounter{table}{0}
\newpage
\section{Data curation}
\begin{figure}[H]
    \centering
    \includegraphics[width=0.95\textwidth]{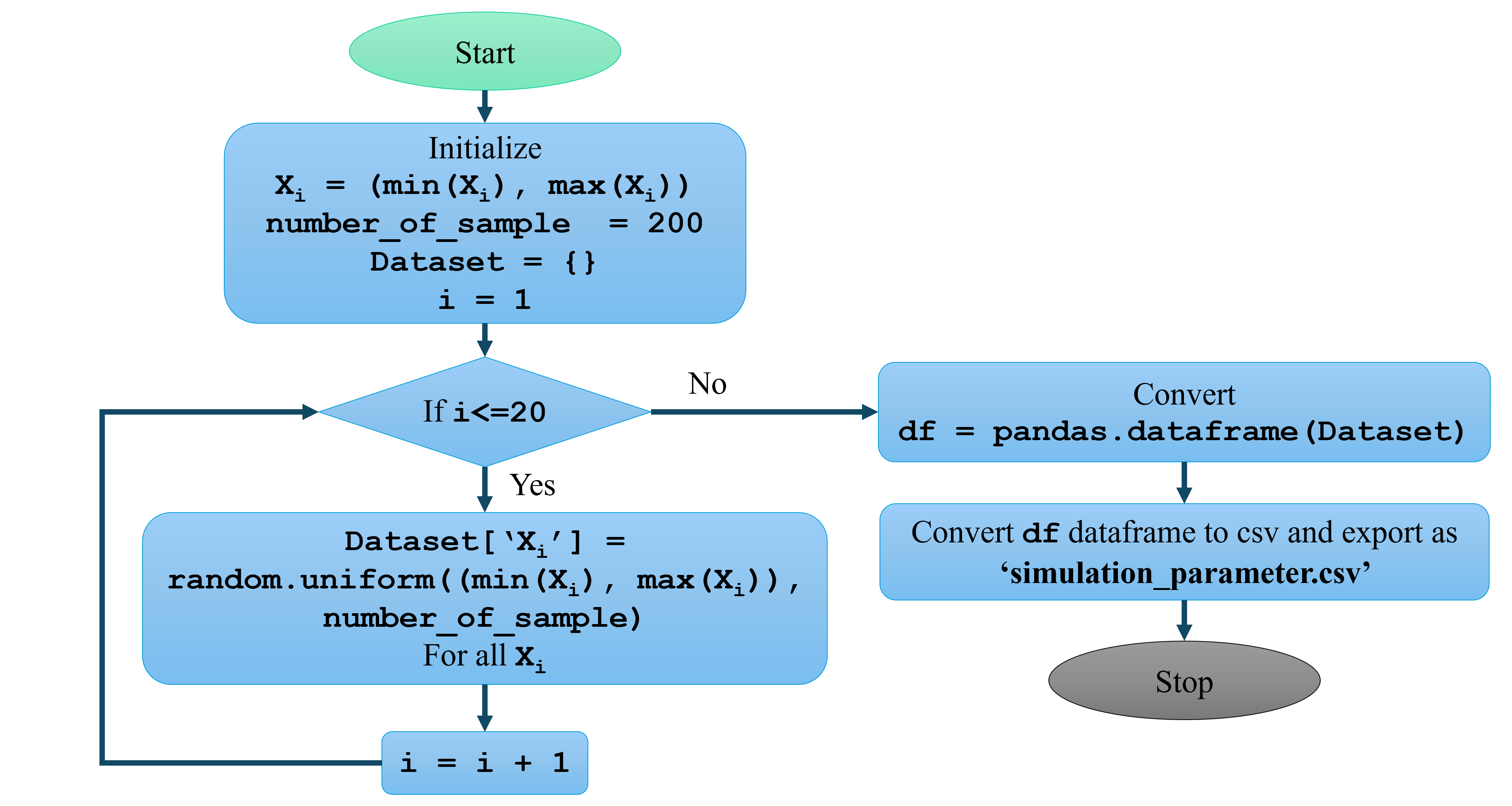}
    \caption{Flowchart illustrating the data generation process for simulation parameter sampling.~\Cref{tab:simulation_parameters_and_their_ranges,tab:feature_mapping} lists all simulation parameters and their upper and lower limits used to generate the dataset, respectively.}
    \label{fig:data_generation_flowchart}
\end{figure}

\noindent \Cref{fig:data_generation_flowchart} illustrates the flowchart for the dataset generation process for a subset of material parameters and geometrical configurations.

\subsection{DEM simulation setup: EDEMpy}
\noindent The generated parameter sets are programmatically transformed into simulation configurations using \texttt{EDEMpy}, a Python-based interface for the DEM solver \texttt{EDEM}. This automation facilitated batch processing of numerous simulations without manual intervention. The simulation setup involved several key steps. A base simulation template is created and systematically copied for each parameter set, allowing consistent file management. Geometrical configuration is performed by computing appropriate rotation matrices and translations for the plate and the particle factory based on the specified angles. The material properties generated earlier are then applied to the particles and equipment surfaces. The Hertz-Mindlin contact model is implemented, incorporating suitable damping and friction parameters to capture inter-particle interactions accurately. The particle factory setup defined the particle generation rate, size distribution, and initial velocity conditions. Archard's wear model is integrated by dynamically specifying wear constants for each configuration. Finally, simulation boundaries are adapted to align with varying geometrical configurations. The \texttt{EDEMpy} implementation used the \texttt{Deck} class to modify simulation files directly, offering fine-grained control over simulation parameters while ensuring uniformity across all generated cases.

The \texttt{EDEMpy} implementation utilized the Deck class to modify simulation files directly, enabling precise parametric control while maintaining consistent simulation conditions across the parameter space. The workflow illustrates the core functionality:
\begin{figure}[H]
    \centering
    \includegraphics[width=0.5\textwidth]{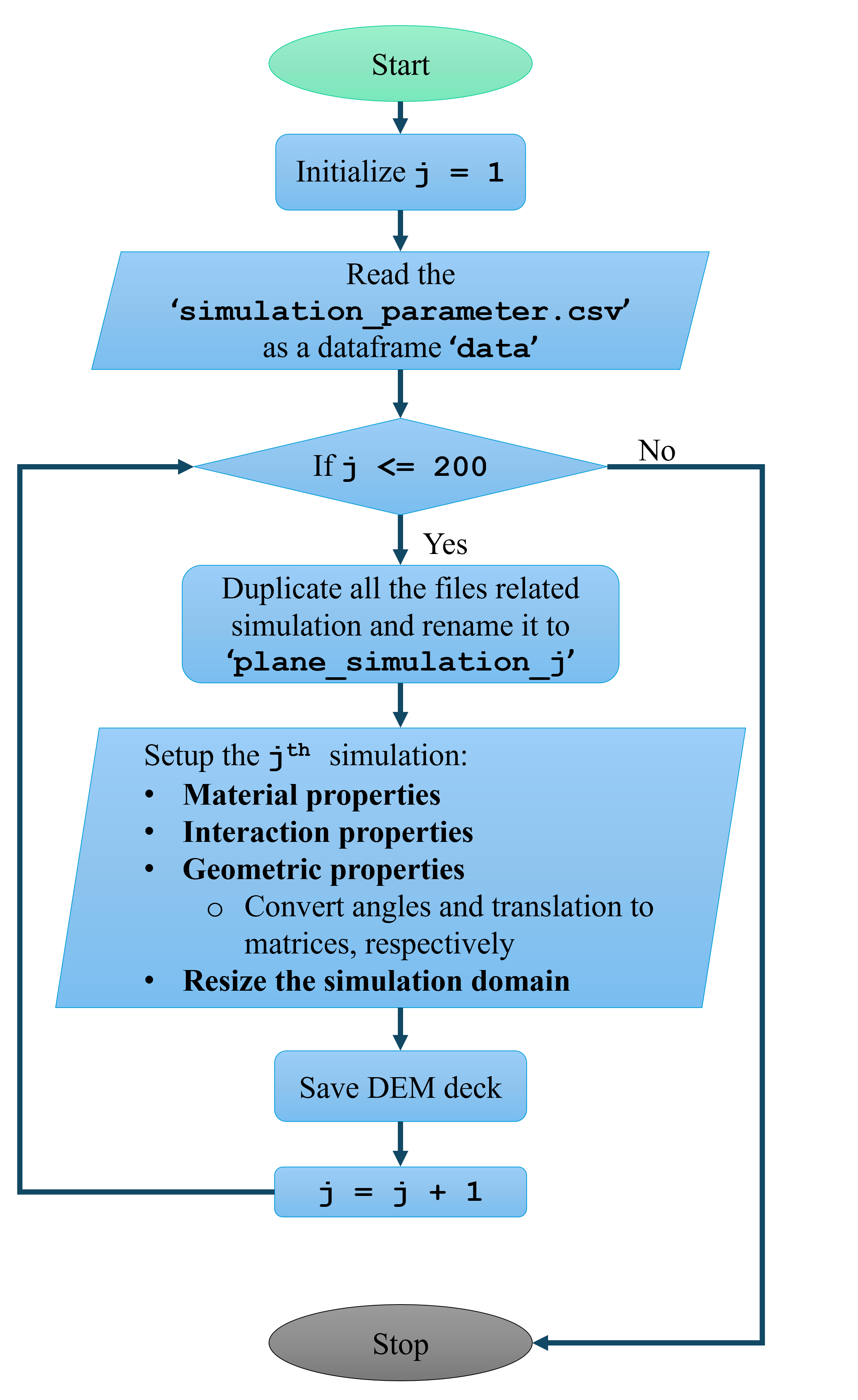}
    \caption{A flowchart illustrating the automated DEM simulation setup.}
    \label{fig:simulation_setup_flowchart}
\end{figure}

\subsection{Batch simulation and data collection}
\noindent A custom Python script is used to generate Windows batch files (.bat), each containing the necessary command-line instructions to run individual simulations with the appropriate parameter configurations. These batch files are executed sequentially on the available computational resources, enabling fully automated simulation workflows. Each simulation is pre-configured to export key data outputs, including wear depth, at regular intervals. Simulation outputs are systematically named and organized within a hierarchical directory to support streamlined post-processing. This automated batch processing approach significantly reduced manual effort and ensured reproducibility, allowing all 200 simulations to be executed efficiently. Each simulation is run until a steady-state wear condition is achieved, typically corresponding to approximately \SI{5}{\second} of simulated time.

\section{Movies}
\begin{itemize}[labelindent=1em,labelsep=0.5cm,leftmargin=*]
    \item \href{https://youtu.be/cT2ZtLeix4w}{Movie 1}: A comparative analysis of wear with 60$^\circ$, 40$^\circ$ and 30$^\circ$ plate angle with horizontal respectively from left to right. The increase in wear can be described as a yellow-to-orange shift in hue, where orange is the maximum wear. 
   
\end{itemize}

\begin{thebibliography}{62}
\expandafter\ifx\csname natexlab\endcsname\relax\def\natexlab#1{#1}\fi
\providecommand{\url}[1]{\texttt{#1}}
\providecommand{\href}[2]{#2}
\providecommand{\path}[1]{#1}
\providecommand{\DOIprefix}{doi:}
\providecommand{\ArXivprefix}{arXiv:}
\providecommand{\URLprefix}{URL: }
\providecommand{\Pubmedprefix}{pmid:}
\providecommand{\doi}[1]{\href{http://dx.doi.org/#1}{\path{#1}}}
\providecommand{\Pubmed}[1]{\href{pmid:#1}{\path{#1}}}
\providecommand{\bibinfo}[2]{#2}
\ifx\xfnm\relax \def\xfnm[#1]{\unskip,\space#1}\fi
\bibitem[{Carter et~al.(1980)Carter, Nobes, and Arshak}]{carter1980mechanism}
\bibinfo{author}{G.~Carter}, \bibinfo{author}{M.~Nobes}, \bibinfo{author}{K.~Arshak},
\newblock \bibinfo{title}{The mechanism of ripple generation on sandblasted ductile solids},
\newblock \bibinfo{journal}{Wear} \bibinfo{volume}{65} (\bibinfo{year}{1980}) \bibinfo{pages}{151--174}.
\bibitem[{Hutchings(1992)}]{Hutchings1992}
\bibinfo{author}{I.~M. Hutchings},
\newblock \bibinfo{title}{Tribology: Friction and wear of engineering materials},
\newblock \bibinfo{journal}{Materials Science and Technology} \bibinfo{volume}{8} (\bibinfo{year}{1992}) \bibinfo{pages}{879--884}.
\bibitem[{Owen and Cleary(2009)}]{owen2009prediction}
\bibinfo{author}{P.~Owen}, \bibinfo{author}{P.~Cleary},
\newblock \bibinfo{title}{Prediction of screw conveyor performance using the discrete element method (dem)},
\newblock \bibinfo{journal}{Powder Technology} \bibinfo{volume}{193} (\bibinfo{year}{2009}) \bibinfo{pages}{274--288}.
\bibitem[{Che et~al.(2025)Che, Peng, Wang, and Wang}]{che2025novel}
\bibinfo{author}{Z.~Che}, \bibinfo{author}{C.~Peng}, \bibinfo{author}{C.~Wang}, \bibinfo{author}{J.~Wang},
\newblock \bibinfo{title}{A novel integrated tdlavoa-xgboost model for tool wear prediction in lathe and milling operations},
\newblock \bibinfo{journal}{Results in Engineering}  (\bibinfo{year}{2025}) \bibinfo{pages}{105984}.
\bibitem[{Suh and Saka(1980)}]{suh1980effect}
\bibinfo{author}{N.~P. Suh}, \bibinfo{author}{N.~Saka},
\newblock \bibinfo{title}{Effect of wear on performance and reliability},
\newblock in: \bibinfo{booktitle}{Risk and Failure Analysis for Improved Performance and Reliability}, \bibinfo{publisher}{Springer}, \bibinfo{year}{1980}, pp. \bibinfo{pages}{243--261}.
\bibitem[{Johanson and Royal(1982)}]{Johanson1982}
\bibinfo{author}{J.~R. Johanson}, \bibinfo{author}{T.~Royal},
\newblock \bibinfo{title}{Measuring and use of wear properties for predicting life of bulk materials handling equipment},
\newblock \bibinfo{journal}{Bulk solids handling} \bibinfo{volume}{2} (\bibinfo{year}{1982}) \bibinfo{pages}{517--523}.
\bibitem[{Graff(2010)}]{graff2010discrete}
\bibinfo{author}{L.~Graff}, \bibinfo{title}{Discrete element method simulation of wear due to soil-tool interaction}, Master's thesis, University of Saskatchewan, \bibinfo{year}{2010}.
\bibitem[{Xia et~al.(2019)Xia, Wang, Li, Wei, and Yang}]{xia2019discrete}
\bibinfo{author}{R.~Xia}, \bibinfo{author}{X.~Wang}, \bibinfo{author}{B.~Li}, \bibinfo{author}{X.~Wei}, \bibinfo{author}{Z.~Yang},
\newblock \bibinfo{title}{Discrete element method-(dem-) based study on the wear mechanism and wear regularity in scraper conveyor chutes},
\newblock \bibinfo{journal}{Mathematical Problems in Engineering} \bibinfo{volume}{2019} (\bibinfo{year}{2019}) \bibinfo{pages}{4191570}.
\bibitem[{Amadi et~al.(2024)Amadi, Mohyaldinn, and Ridha}]{Amadi2024}
\bibinfo{author}{A.~Amadi}, \bibinfo{author}{M.~Mohyaldinn}, \bibinfo{author}{S.~Ridha},
\newblock \bibinfo{title}{Sand particle induced erosion in oil and gas screens: A review of influencing factors and wear dynamics},
\newblock \bibinfo{journal}{Powder Technology} \bibinfo{volume}{436} (\bibinfo{year}{2024}) \bibinfo{pages}{119528}.
\bibitem[{Fillot et~al.(2007)Fillot, Iordanoff, and Berthier}]{fillot2007modelling}
\bibinfo{author}{N.~Fillot}, \bibinfo{author}{I.~Iordanoff}, \bibinfo{author}{Y.~Berthier},
\newblock \bibinfo{title}{Modelling third body flows with a discrete element method—a tool for understanding wear with adhesive particles},
\newblock \bibinfo{journal}{Tribology International} \bibinfo{volume}{40} (\bibinfo{year}{2007}) \bibinfo{pages}{973--981}.
\bibitem[{Jerier and Molinari(2012)}]{jerier2012normal}
\bibinfo{author}{J.~Jerier}, \bibinfo{author}{J.~Molinari},
\newblock \bibinfo{title}{Normal contact between rough surfaces by the discrete element method},
\newblock \bibinfo{journal}{Tribology International} \bibinfo{volume}{47} (\bibinfo{year}{2012}) \bibinfo{pages}{1--8}.
\bibitem[{Pozzetti and Peters(2018)}]{pozzetti2018numerical}
\bibinfo{author}{G.~Pozzetti}, \bibinfo{author}{B.~Peters},
\newblock \bibinfo{title}{A numerical approach for the evaluation of particle-induced erosion in an abrasive waterjet focusing tube},
\newblock \bibinfo{journal}{Powder Technology} \bibinfo{volume}{333} (\bibinfo{year}{2018}) \bibinfo{pages}{229--242}.
\bibitem[{Fransen et~al.(2021)Fransen, Langelaar, and Schott}]{fransen2021application}
\bibinfo{author}{M.~P. Fransen}, \bibinfo{author}{M.~Langelaar}, \bibinfo{author}{D.~L. Schott},
\newblock \bibinfo{title}{Application of dem-based metamodels in bulk handling equipment design: Methodology and dem case study},
\newblock \bibinfo{journal}{Powder Technology} \bibinfo{volume}{393} (\bibinfo{year}{2021}) \bibinfo{pages}{205--218}.
\bibitem[{Thompson et~al.(2022)Thompson, Berry, Southern, Walls, Holmes, and Brown}]{thompson2022effect}
\bibinfo{author}{J.~A. Thompson}, \bibinfo{author}{L.~Berry}, \bibinfo{author}{S.~Southern}, \bibinfo{author}{W.~K. Walls}, \bibinfo{author}{M.~A. Holmes}, \bibinfo{author}{S.~G. Brown},
\newblock \bibinfo{title}{The effect of mesh discretisation on damage and wear predictions using the discrete element method},
\newblock \bibinfo{journal}{Applied Mathematical Modelling} \bibinfo{volume}{105} (\bibinfo{year}{2022}) \bibinfo{pages}{690--710}.
\bibitem[{Archard(1953)}]{Archard1953}
\bibinfo{author}{J.~F. Archard},
\newblock \bibinfo{title}{Contact and rubbing of flat surfaces},
\newblock \bibinfo{journal}{Journal of Applied Physics} \bibinfo{volume}{24} (\bibinfo{year}{1953}) \bibinfo{pages}{981--988}.
\bibitem[{Jayasundara and Zhu(2022)}]{jayasundara2022predicting}
\bibinfo{author}{C.~Jayasundara}, \bibinfo{author}{H.~Zhu},
\newblock \bibinfo{title}{Predicting liner wear of ball mills using discrete element method and artificial neural network},
\newblock \bibinfo{journal}{Chemical Engineering Research and Design} \bibinfo{volume}{182} (\bibinfo{year}{2022}) \bibinfo{pages}{438--447}.
\bibitem[{Wang et~al.(2023)Wang, Liu, Tong, Xu, and Ma}]{wang2023parameter}
\bibinfo{author}{S.~Wang}, \bibinfo{author}{X.~Liu}, \bibinfo{author}{T.~Tong}, \bibinfo{author}{Z.~Xu}, \bibinfo{author}{Y.~Ma},
\newblock \bibinfo{title}{Parameter optimization and dem simulation of bionic sweep with lower abrasive wear characteristics},
\newblock \bibinfo{journal}{Biomimetics} \bibinfo{volume}{8} (\bibinfo{year}{2023}) \bibinfo{pages}{201}.
\bibitem[{Yan et~al.(2023)Yan, Helmons, Carr, Wheeler, and Schott}]{yan2023modelling}
\bibinfo{author}{Y.~Yan}, \bibinfo{author}{R.~Helmons}, \bibinfo{author}{M.~Carr}, \bibinfo{author}{C.~Wheeler}, \bibinfo{author}{D.~Schott},
\newblock \bibinfo{title}{Modelling of material removal due to sliding wear caused by bulk material},
\newblock \bibinfo{journal}{Powder Technology} \bibinfo{volume}{415} (\bibinfo{year}{2023}) \bibinfo{pages}{118109}.
\bibitem[{Liskiewicz et~al.(2023)Liskiewicz, Sherrington, Khan, and Liu}]{liskiewicz2023advances}
\bibinfo{author}{T.~Liskiewicz}, \bibinfo{author}{I.~Sherrington}, \bibinfo{author}{T.~Khan}, \bibinfo{author}{Y.~Liu},
\newblock \bibinfo{title}{Advances in sensing for real-time monitoring of tribological parameters},
\newblock \bibinfo{journal}{Tribology International} \bibinfo{volume}{189} (\bibinfo{year}{2023}) \bibinfo{pages}{108965}.
\bibitem[{Zhang and Li(2024)}]{Zhang2024}
\bibinfo{author}{Y.~Zhang}, \bibinfo{author}{X.~Li},
\newblock \bibinfo{title}{Intelligent tool wear prediction based on deep learning psd-cvt},
\newblock \bibinfo{journal}{Scientific Reports} \bibinfo{volume}{14} (\bibinfo{year}{2024}) \bibinfo{pages}{12345}.
\bibitem[{Yan et~al.(2015)Yan, Wilkinson, Stitt, and Marigo}]{yan2015discrete}
\bibinfo{author}{Z.~Yan}, \bibinfo{author}{S.~Wilkinson}, \bibinfo{author}{E.~Stitt}, \bibinfo{author}{M.~Marigo},
\newblock \bibinfo{title}{Discrete element modelling (dem) input parameters: understanding their impact on model predictions using statistical analysis},
\newblock \bibinfo{journal}{Computational Particle Mechanics} \bibinfo{volume}{2} (\bibinfo{year}{2015}) \bibinfo{pages}{283--299}.
\bibitem[{Wallin and Servin(2022)}]{wallin2022data}
\bibinfo{author}{E.~Wallin}, \bibinfo{author}{M.~Servin},
\newblock \bibinfo{title}{Data-driven model order reduction for granular media},
\newblock \bibinfo{journal}{Computational Particle Mechanics} \bibinfo{volume}{9} (\bibinfo{year}{2022}) \bibinfo{pages}{15--28}.
\bibitem[{Iraz{\'a}bal et~al.(2023)Iraz{\'a}bal, Salazar, and Vicente}]{irazabal2023methodology}
\bibinfo{author}{J.~Iraz{\'a}bal}, \bibinfo{author}{F.~Salazar}, \bibinfo{author}{D.~J. Vicente},
\newblock \bibinfo{title}{A methodology for calibrating parameters in discrete element models based on machine learning surrogates},
\newblock \bibinfo{journal}{Computational Particle Mechanics} \bibinfo{volume}{10} (\bibinfo{year}{2023}) \bibinfo{pages}{1031--1047}.
\bibitem[{Jin et~al.(2023)Jin, Zhang, and Espinosa}]{jin2023recent}
\bibinfo{author}{H.~Jin}, \bibinfo{author}{E.~Zhang}, \bibinfo{author}{H.~D. Espinosa},
\newblock \bibinfo{title}{Recent advances and applications of machine learning in experimental solid mechanics: A review},
\newblock \bibinfo{journal}{Applied Mechanics Reviews} \bibinfo{volume}{75} (\bibinfo{year}{2023}) \bibinfo{pages}{061001}.
\bibitem[{Sose et~al.(2023)Sose, Joshi, Kunche, Wang, and Deshmukh}]{sose2023review}
\bibinfo{author}{A.~T. Sose}, \bibinfo{author}{S.~Y. Joshi}, \bibinfo{author}{L.~K. Kunche}, \bibinfo{author}{F.~Wang}, \bibinfo{author}{S.~A. Deshmukh},
\newblock \bibinfo{title}{A review of recent advances and applications of machine learning in tribology},
\newblock \bibinfo{journal}{Physical Chemistry Chemical Physics} \bibinfo{volume}{25} (\bibinfo{year}{2023}) \bibinfo{pages}{4408--4443}.
\bibitem[{Wang et~al.(2025)Wang, Kumar, Feng, Qu, and Wang}]{wang2025machine}
\bibinfo{author}{M.~Wang}, \bibinfo{author}{K.~Kumar}, \bibinfo{author}{Y.~Feng}, \bibinfo{author}{T.~Qu}, \bibinfo{author}{M.~Wang},
\newblock \bibinfo{title}{Machine learning aided modeling of granular materials: A review: M. wang et al.},
\newblock \bibinfo{journal}{Archives of Computational Methods in Engineering} \bibinfo{volume}{32} (\bibinfo{year}{2025}) \bibinfo{pages}{1997--2034}.
\bibitem[{Rajput and Das(2023)}]{Rajput2023}
\bibinfo{author}{A.~S. Rajput}, \bibinfo{author}{S.~Das},
\newblock \bibinfo{title}{A machine learning approach to predict the wear behaviour of steels},
\newblock \bibinfo{journal}{Tribology International} \bibinfo{volume}{185} (\bibinfo{year}{2023}) \bibinfo{pages}{108500}.
\bibitem[{Fathi et~al.(2024)Fathi, Chen, Abdallah, and Saleh}]{Fathi2024}
\bibinfo{author}{R.~Fathi}, \bibinfo{author}{M.~Chen}, \bibinfo{author}{M.~Abdallah}, \bibinfo{author}{B.~Saleh},
\newblock \bibinfo{title}{Wear prediction of functionally graded composites using machine learning},
\newblock \bibinfo{journal}{Materials} \bibinfo{volume}{17} (\bibinfo{year}{2024}) \bibinfo{pages}{4523}.
\bibitem[{Danish et~al.(2024)Danish, Gupta, Irfan, Ghazali, Rathore, Krolczyk, and Alsaady}]{Danish2024}
\bibinfo{author}{M.~Danish}, \bibinfo{author}{M.~K. Gupta}, \bibinfo{author}{S.~A. Irfan}, \bibinfo{author}{S.~M. Ghazali}, \bibinfo{author}{M.~F. Rathore}, \bibinfo{author}{G.~M. Krolczyk}, \bibinfo{author}{A.~Alsaady},
\newblock \bibinfo{title}{Machine learning models for prediction and classification of tool wear in sustainable milling of additively manufactured 316 stainless steel},
\newblock \bibinfo{journal}{Results in Engineering} \bibinfo{volume}{22} (\bibinfo{year}{2024}) \bibinfo{pages}{102015}.
\bibitem[{Hussain et~al.(2024)Hussain, Sakhaei, and Shafiee}]{hussain2024machine}
\bibinfo{author}{A.~Hussain}, \bibinfo{author}{A.~H. Sakhaei}, \bibinfo{author}{M.~Shafiee},
\newblock \bibinfo{title}{Machine learning-based constitutive modelling for material non-linearity: A review},
\newblock \bibinfo{journal}{Mechanics of Advanced Materials and Structures}  (\bibinfo{year}{2024}) \bibinfo{pages}{1--19}.
\bibitem[{Hasan and Nosonovsky(2022)}]{Hasan2022}
\bibinfo{author}{M.~S. Hasan}, \bibinfo{author}{M.~Nosonovsky},
\newblock \bibinfo{title}{Triboinformatics: Machine learning algorithms and data topology methods for tribology},
\newblock \bibinfo{journal}{Surface Innovations} \bibinfo{volume}{10} (\bibinfo{year}{2022}) \bibinfo{pages}{229--242}.
\bibitem[{Zhu et~al.(2024)Zhu, Jin, Li, Han, and Yan}]{Zhu2024}
\bibinfo{author}{C.~Zhu}, \bibinfo{author}{L.~Jin}, \bibinfo{author}{W.~Li}, \bibinfo{author}{S.~Han}, \bibinfo{author}{J.~Yan},
\newblock \bibinfo{title}{The prediction of wear depth based on machine learning algorithms},
\newblock \bibinfo{journal}{Lubricants} \bibinfo{volume}{12} (\bibinfo{year}{2024}) \bibinfo{pages}{34}.
\bibitem[{Altair-Engineering(2023)}]{Engineering2023}
\bibinfo{author}{Altair-Engineering}, \bibinfo{title}{The archard wear model}, \bibinfo{howpublished}{EDEM Documentation}, \bibinfo{year}{2023}.
\bibitem[{Lommen et~al.(2019)Lommen, Mohajeri, Lodewijks, and Schott}]{lommen2019particle}
\bibinfo{author}{S.~Lommen}, \bibinfo{author}{M.~Mohajeri}, \bibinfo{author}{G.~Lodewijks}, \bibinfo{author}{D.~Schott},
\newblock \bibinfo{title}{Dem particle upscaling for large-scale bulk handling equipment and material interaction},
\newblock \bibinfo{journal}{Powder Technology} \bibinfo{volume}{352} (\bibinfo{year}{2019}) \bibinfo{pages}{273--282}.
\bibitem[{Deshpande et~al.(2024)Deshpande, Kulkarni, Wasatkar, Gajalkar, and Abdullah}]{Deshpande2024}
\bibinfo{author}{A.~R. Deshpande}, \bibinfo{author}{A.~P. Kulkarni}, \bibinfo{author}{N.~Wasatkar}, \bibinfo{author}{V.~Gajalkar}, \bibinfo{author}{M.~Abdullah},
\newblock \bibinfo{title}{Prediction of wear rate of glass-filled ptfe composites based on machine learning approaches},
\newblock \bibinfo{journal}{Polymers} \bibinfo{volume}{16} (\bibinfo{year}{2024}) \bibinfo{pages}{2666}.
\bibitem[{Yan et~al.(2023)Yan, Helmons, Carr, Wheeler, and Schott}]{Yan2023}
\bibinfo{author}{Y.~Yan}, \bibinfo{author}{R.~Helmons}, \bibinfo{author}{M.~Carr}, \bibinfo{author}{C.~Wheeler}, \bibinfo{author}{D.~Schott},
\newblock \bibinfo{title}{Modelling of material removal due to sliding wear caused by bulk material},
\newblock \bibinfo{journal}{Powder Technology} \bibinfo{volume}{415} (\bibinfo{year}{2023}) \bibinfo{pages}{118109}.
\bibitem[{Coetzee(2017)}]{coetzee2017calibration}
\bibinfo{author}{C.~J. Coetzee},
\newblock \bibinfo{title}{Calibration of the discrete element method},
\newblock \bibinfo{journal}{Powder Technology} \bibinfo{volume}{310} (\bibinfo{year}{2017}) \bibinfo{pages}{104--142}.
\bibitem[{Mindlin and Deresiewicz(1953)}]{mindlin1953elastic}
\bibinfo{author}{R.~D. Mindlin}, \bibinfo{author}{H.~Deresiewicz},
\newblock \bibinfo{title}{Elastic spheres in contact under varying oblique forces},
\newblock \bibinfo{journal}{JOurnal of Applied Mechanics}  (\bibinfo{year}{1953}).
\bibitem[{Johnson(1987)}]{johnson1987contact}
\bibinfo{author}{K.~L. Johnson}, \bibinfo{title}{Contact mechanics}, \bibinfo{publisher}{Cambridge University Press}, \bibinfo{year}{1987}.
\bibitem[{Thornton et~al.(2011)Thornton, Cummins, and Cleary}]{thornton2011investigation}
\bibinfo{author}{C.~Thornton}, \bibinfo{author}{S.~J. Cummins}, \bibinfo{author}{P.~W. Cleary},
\newblock \bibinfo{title}{An investigation of the comparative behaviour of alternative contact force models during elastic collisions},
\newblock \bibinfo{journal}{Powder Technology} \bibinfo{volume}{210} (\bibinfo{year}{2011}) \bibinfo{pages}{189--197}.
\bibitem[{Rojas et~al.(2019)Rojas, Vergara, and Soto}]{rojas2019case}
\bibinfo{author}{E.~Rojas}, \bibinfo{author}{V.~Vergara}, \bibinfo{author}{R.~Soto},
\newblock \bibinfo{title}{Case study: Discrete element modeling of wear in mining hoppers},
\newblock \bibinfo{journal}{Wear} \bibinfo{volume}{430} (\bibinfo{year}{2019}) \bibinfo{pages}{120--125}.
\bibitem[{Forsstr{\"o}m and Jons{\'e}n(2016)}]{forsstrom2016calibration}
\bibinfo{author}{D.~Forsstr{\"o}m}, \bibinfo{author}{P.~Jons{\'e}n},
\newblock \bibinfo{title}{Calibration and validation of a large scale abrasive wear model by coupling dem-fem: Local failure prediction from abrasive wear of tipper bodies during unloading of granular material},
\newblock \bibinfo{journal}{Engineering Failure Analysis} \bibinfo{volume}{66} (\bibinfo{year}{2016}) \bibinfo{pages}{274--283}.
\bibitem[{Chen et~al.(2017)Chen, Schott, and Lodewijks}]{chen2017sensitivity}
\bibinfo{author}{G.~Chen}, \bibinfo{author}{D.~L. Schott}, \bibinfo{author}{G.~Lodewijks},
\newblock \bibinfo{title}{Sensitivity analysis of dem prediction for sliding wear by single iron ore particle},
\newblock \bibinfo{journal}{Engineering Computations} \bibinfo{volume}{34} (\bibinfo{year}{2017}) \bibinfo{pages}{2031--2053}.
\bibitem[{Aalen(1989)}]{aalen1989linear}
\bibinfo{author}{O.~O. Aalen},
\newblock \bibinfo{title}{A linear regression model for the analysis of life times},
\newblock \bibinfo{journal}{Statistics in medicine} \bibinfo{volume}{8} (\bibinfo{year}{1989}) \bibinfo{pages}{907--925}.
\bibitem[{Montgomery et~al.(2012)Montgomery, Peck, and Vining}]{montgomery2012introduction}
\bibinfo{author}{D.~C. Montgomery}, \bibinfo{author}{E.~A. Peck}, \bibinfo{author}{G.~G. Vining}, \bibinfo{title}{Introduction to Linear Regression Analysis}, \bibinfo{edition}{5th} ed., \bibinfo{publisher}{John Wiley \& Sons}, \bibinfo{year}{2012}.
\bibitem[{Tibshirani(1996)}]{tibshirani1996regression}
\bibinfo{author}{R.~Tibshirani},
\newblock \bibinfo{title}{Regression shrinkage and selection via the lasso},
\newblock \bibinfo{journal}{Journal of the Royal Statistical Society: Series B (Methodological)} \bibinfo{volume}{58} (\bibinfo{year}{1996}) \bibinfo{pages}{267--288}.
\bibitem[{Santosa and Symes(1986)}]{santosa1986linear}
\bibinfo{author}{F.~Santosa}, \bibinfo{author}{W.~W. Symes},
\newblock \bibinfo{title}{Linear inversion of band-limited reflection seismograms},
\newblock \bibinfo{journal}{SIAM journal on scientific and statistical computing} \bibinfo{volume}{7} (\bibinfo{year}{1986}) \bibinfo{pages}{1307--1330}.
\bibitem[{Hayashi(1998)}]{hayashi1998data}
\bibinfo{author}{C.~Hayashi},
\newblock \bibinfo{title}{What is data science? fundamental concepts and a heuristic example},
\newblock in: \bibinfo{booktitle}{Data Science, Classification, and Related Methods: Proceedings of the Fifth Conference of the International Federation of Classification Societies (IFCS-96), Kobe, Japan, March 27--30, 1996}, \bibinfo{organization}{Springer}, \bibinfo{year}{1998}, pp. \bibinfo{pages}{40--51}.
\bibitem[{Hoerl and Kennard(1970)}]{hoerl1970ridge}
\bibinfo{author}{A.~E. Hoerl}, \bibinfo{author}{R.~W. Kennard},
\newblock \bibinfo{title}{Ridge regression: Biased estimation for nonorthogonal problems},
\newblock \bibinfo{journal}{Technometrics} \bibinfo{volume}{12} (\bibinfo{year}{1970}) \bibinfo{pages}{55--67}.
\bibitem[{Herawati et~al.(2024)Herawati, Sutrisno, Nusyirwan, Misgiyati et~al.}]{herawati2024performance}
\bibinfo{author}{N.~Herawati}, \bibinfo{author}{A.~Sutrisno}, \bibinfo{author}{N.~Nusyirwan}, \bibinfo{author}{M.~Misgiyati}, et~al.,
\newblock \bibinfo{title}{The performance of ridge regression, lasso, and elastic-net in controlling multicollinearity: A simulation and application},
\newblock \bibinfo{journal}{Journal of Modern Applied Statistical Methods,} \bibinfo{volume}{23} (\bibinfo{year}{2024}) \bibinfo{pages}{1--13}.
\bibitem[{Jolliffe and Cadima(2016)}]{JolliffeCadima2016}
\bibinfo{author}{I.~T. Jolliffe}, \bibinfo{author}{J.~Cadima},
\newblock \bibinfo{title}{Principal component analysis: a review and recent developments},
\newblock \bibinfo{journal}{Philosophical Transactions of the Royal Society A: Mathematical, Physical and Engineering Sciences} \bibinfo{volume}{374} (\bibinfo{year}{2016}) \bibinfo{pages}{20150202}.
\bibitem[{Arg{\"u}elles et~al.(2014)Arg{\"u}elles, Benavides, and Fern{\'a}ndez}]{arguelles2014new}
\bibinfo{author}{M.~Arg{\"u}elles}, \bibinfo{author}{C.~Benavides}, \bibinfo{author}{I.~Fern{\'a}ndez},
\newblock \bibinfo{title}{A new approach to the identification of regional clusters: hierarchical clustering on principal components},
\newblock \bibinfo{journal}{Applied Economics} \bibinfo{volume}{46} (\bibinfo{year}{2014}) \bibinfo{pages}{2511--2519}.
\bibitem[{Ibrahim et~al.(2021)Ibrahim, Nazir, and Velastin}]{ibrahim2021feature}
\bibinfo{author}{S.~Ibrahim}, \bibinfo{author}{S.~Nazir}, \bibinfo{author}{S.~A. Velastin},
\newblock \bibinfo{title}{Feature selection using correlation analysis and principal component analysis for accurate breast cancer diagnosis},
\newblock \bibinfo{journal}{Journal of imaging} \bibinfo{volume}{7} (\bibinfo{year}{2021}) \bibinfo{pages}{225}.
\bibitem[{Breiman et~al.(1984)Breiman, Friedman, Olshen, and Stone}]{Breiman1984}
\bibinfo{author}{L.~Breiman}, \bibinfo{author}{J.~Friedman}, \bibinfo{author}{R.~Olshen}, \bibinfo{author}{C.~Stone}, \bibinfo{title}{Classification and Regression Trees}, \bibinfo{publisher}{Chapman and Hall/CRC}, \bibinfo{year}{1984}.
\bibitem[{Boehmke and Greenwell(2019)}]{Bohemke2020}
\bibinfo{author}{B.~Boehmke}, \bibinfo{author}{B.~M. Greenwell}, \bibinfo{title}{Hands-on machine learning with R}, \bibinfo{publisher}{Chapman and Hall/CRC}, \bibinfo{year}{2019}.
\bibitem[{Loh(2014)}]{Loh2014}
\bibinfo{author}{W.~Loh},
\newblock \bibinfo{title}{Fifty years of classification and regression trees},
\newblock \bibinfo{journal}{International Statistical Review} \bibinfo{volume}{82} (\bibinfo{year}{2014}) \bibinfo{pages}{329--348}.
\bibitem[{Ali et~al.(2012)Ali, Khan, Ahmad, and Maqsood}]{ali2012random}
\bibinfo{author}{J.~Ali}, \bibinfo{author}{R.~Khan}, \bibinfo{author}{N.~Ahmad}, \bibinfo{author}{I.~Maqsood},
\newblock \bibinfo{title}{Random forests and decision trees},
\newblock \bibinfo{journal}{International Journal of Computer Science Issues (IJCSI)} \bibinfo{volume}{9} (\bibinfo{year}{2012}) \bibinfo{pages}{272}.
\bibitem[{Goodfellow et~al.(2016)Goodfellow, Bengio, and Courville}]{Goodfellow2016Deep}
\bibinfo{author}{I.~Goodfellow}, \bibinfo{author}{Y.~Bengio}, \bibinfo{author}{A.~Courville}, \bibinfo{title}{Deep Learning}, \bibinfo{publisher}{MIT Press}, \bibinfo{year}{2016}.
\bibitem[{Yao(1999)}]{Yao1999Evolving}
\bibinfo{author}{X.~Yao},
\newblock \bibinfo{title}{Evolving artificial neural networks},
\newblock \bibinfo{journal}{Proceedings of the IEEE} \bibinfo{volume}{87} (\bibinfo{year}{1999}) \bibinfo{pages}{1423--1447}.
\bibitem[{Nikbakht et~al.(2021)Nikbakht, Anitescu, and Rabczuk}]{Nikbakht2021Optimizing}
\bibinfo{author}{S.~Nikbakht}, \bibinfo{author}{C.~Anitescu}, \bibinfo{author}{T.~Rabczuk},
\newblock \bibinfo{title}{Optimizing the neural network hyperparameters utilizing genetic algorithm},
\newblock \bibinfo{journal}{Journal of Zhejiang University-Science A} \bibinfo{volume}{22} (\bibinfo{year}{2021}) \bibinfo{pages}{407--426}.
\bibitem[{Xiao et~al.(2020)Xiao, Yan, Basodi, Ji, and Pan}]{Xiao2020Efficient}
\bibinfo{author}{X.~Xiao}, \bibinfo{author}{M.~Yan}, \bibinfo{author}{S.~Basodi}, \bibinfo{author}{C.~Ji}, \bibinfo{author}{Y.~Pan},
\newblock \bibinfo{title}{Efficient hyperparameter optimization in deep learning using a variable length genetic algorithm},
\newblock \bibinfo{journal}{arXiv preprint arXiv:2006.12703}  (\bibinfo{year}{2020}).
\bibitem[{Yang et~al.(2021)Yang, Meng, Dai, Yin, Yao, and Yuan}]{yang2021screwwear}
\bibinfo{author}{W.~Yang}, \bibinfo{author}{W.~Meng}, \bibinfo{author}{X.~Dai}, \bibinfo{author}{Z.~Yin}, \bibinfo{author}{F.~Yao}, \bibinfo{author}{Y.~Yuan},
\newblock \bibinfo{title}{Continuous medium hypothesis-based study on the screw flight wear model and wear regularity in a screw ship unloader},
\newblock \bibinfo{journal}{Transactions of the Canadian Society for Mechanical Engineering} \bibinfo{volume}{45} (\bibinfo{year}{2021}) \bibinfo{pages}{584--593}.

\end{thebibliography}
\end{document}